\documentclass[11pt, a4]{article}

\usepackage{amsmath,amssymb,graphicx}
\usepackage{lineno}
\usepackage[usenames,dvipsnames]{color}
\usepackage{subfigure}
\usepackage[margin=1in]{geometry}
\usepackage{yhmath}
\usepackage{ulem}

\newcommand{\be}{\begin{equation}}
\newcommand{\ee}{\end{equation}}
\newcommand{\bea}{\begin{eqnarray}}
\newcommand{\eea}{\end{eqnarray}}
\newcommand{\bes}{\begin{equation*}}
\newcommand{\ees}{\end{equation*}}
\newcommand{\beas}{\begin{eqnarray*}}
\newcommand{\eeas}{\end{eqnarray*}}

\begin{document}
\date{}

  \title{\bf Radiating dispersive shock waves \\ in nonlocal optical media}
  \author{Gennady~A.~El  $^1$ and Noel~F.~Smyth $^2$ \\ \\
$^1$ Department of Mathematical Sciences, Loughborough University, U.K. \\
$^2$ School of Mathematics, University of Edinburgh,  Scotland, U.K.} 

\maketitle


\begin{abstract}
We consider the step Riemann problem for the system of equations describing the 
propagation of a coherent light beam in nematic liquid crystals, which is a general 
system describing nonlinear wave propagation in a number of different physical 
applications.  While the equation governing the light beam is of defocusing nonlinear 
Schr\"odinger equation type, the dispersive shock wave (DSW) generated from this initial condition 
has major differences from the standard DSW solution of the defocusing 
nonlinear Schr\"odinger equation.  In particular, it is found that the DSW has 
positive polarity and generates resonant radiation which propagates ahead of it.  
Remarkably, the velocity of the lead soliton of the DSW is determined by the 
classical shock velocity.  The solution for the radiative wavetrain is obtained using 
the WKB approximation.  It is shown that for sufficiently small initial jumps the nematic 
DSW is asymptotically governed by a Korteweg-de Vries equation with fifth order 
dispersion, which explicitly shows the resonance generating the radiation ahead of the DSW.  
The constructed asymptotic theory is shown to be in good agreement with the results 
of direct numerical simulations. 
\end{abstract}

\section{Introduction}
Dispersive shock waves (DSWs), also termed undular bores in fluid mechanics, are generic solutions of 
nonlinear dispersive wave equations, including the Korteweg-de Vries (KdV), nonlinear Schr\"odinger (NLS) and 
Sine-Gordon equations.  A DSW forms due to the dispersive resolution of a discontinuity and is the 
dispersive equivalent of a gas dynamic shock for which a discontinuity is resolved by viscosity \cite{whitham}.  A DSW
is a non-steady modulated wavetrain which continually expands and has solitary waves at its leading edge and linear,
small amplitude waves at its trailing edge (for the case of negative dispersion; if dispersion is positive then the 
orientation of the DSW, i.e.\ the relative position of the linear and soliton edges, changes).  This 
modulated wavetrain provides an oscillatory transition between the two levels of the initial discontinuity.

DSWs/undular bores are a common wave form which can be found in a broad array of physical systems.  The classical 
undular bore is the tidal bore found in regions of large tidal flows and suitable topography, for example the 
Severn Estuary in England and the Bay of Fundy in Canada.  However, undular bores arise in a wide range of 
fluid systems, including the atmosphere, an example being morning glory clouds \cite{clarke,anne}, and the 
semi-diurnal internal tide \cite{nwshelf}.
They also arise in geophysics (magma flow) \cite{scott1,magma,hoefer2} and Fermi gases \cite{hoefer1}.  
Of particular relevance to the present work, they arise in nonlinear optics for a wide range of 
optical materials, including photorefractive crystals \cite{fleischer, fleischer2,elopt}, optical fibres 
\cite{trillo1,trillo7,trillores,trilloresfour,trilloreslossbore,trilloresnature}, nonlinear thermal
optical media \cite{trillo6,boreexp}, colloidal media \cite{focbore2,boreapprox} 
and nematic liquid crystals \cite{boreapprox,nembore}.  

DSW solutions of nonlinear dispersive wave equations are usually found using Whitham modulation theory 
\cite{mod1,modproc, whitham}.  Whitham modulation theory is a method for analysing slowly varying (modulated) 
wavetrains and deriving equations for the parameters, mean height, wavenumber, amplitude, etc., of such wavetrains.  
It is equivalent to the method of multiple scales, but much simpler than this to implement.  When the 
underlying wavetrain is stable, the modulation equations form a hyperbolic system for the wavetrain parameters.  
It was found that a simple wave solution of the hyperbolic modulation equations for the KdV equation corresponds 
to a DSW \cite{gur,bengt}, so that the standard method for finding DSW
solutions is from the modulation equations for the relevant governing equation.  This 
original method due to Gurevich and Pitaevskii \cite{gur} and Fornberg and Whitham 
\cite{bengt} relies on the hyperbolic modulation equations being in Riemann invariant 
form, which is guaranteed if the governing equation is integrable with an inverse 
scattering solution \cite{flash}.  However, most equations governing DSWs in 
physical applications are not integrable.  This limitation was overcome to a certain 
extent when it was found that the leading, soliton, edge and trailing, small amplitude 
wave, edge of a DSW could be determined without a knowledge of the full Whitham 
modulation equations \cite{el2}.  

In the present work a DSW due to coherent light propagation in a nematic liquid crystal is analysed.  
While the specific context is light propagation in a nematic liquid crystal, equations similar to those for 
light propagation in this medium also arise for other nonlinear optical media 
\cite{lead,kuz,alejandro1,photo,segev,segev2,trillo6}, in fluid mechanics \cite{alpha1} and in 
models of quantum gravity \cite{grav}.  An optical DSW in a nematic liquid crystal is found to possess 
a number of unique features.  While the equation governing the optical field in a nematic liquid crystal is of 
defocusing NLS-type \cite{gennady}, the DSW is found to be of positive polarity, KdV-type, due to the 
effect of the nematic optical medium, which has a highly ``nonlocal'' response \cite{conti2,PR,Wiley}.  It is 
further found that the dispersion relation for linear waves is non-convex, so that there is a resonance between 
the DSW and dispersive radiation.  This results in a resonant wavetrain propagating ahead of the DSW.  
A similar resonant coupling between a DSW and radiation was found for nonlinear optical beam propagation in
optical fibres when higher order dispersive terms were included in the governing NLS equation to enable such coupling, 
both without \cite{trillores,trilloresfour,trilloresnature} and with \cite{trilloreslossbore} loss.  The driving
mechanism is the resonant coupling with higher order dispersion, which can also occur with just a soliton 
\cite{ressolhnls}.  The total structure of the Riemann problem solution is then found to consist of four distinct 
regions, (i) an expansion wave linking the initial level behind to an intermediate shelf, (ii) a KdV-type 
DSW on this shelf, (iii) a resonant wavetrain leading the DSW and (iv) a front bringing the resonant 
wavetrain down to the initial level ahead.  Asymptotic solutions for all these four regions are 
obtained and compared with full numerical solutions of the governing equations, with generally
excellent agreement being found.  

The paper is organised as follows.  In Section \ref{s:nemeqn} the equations governing light beam propagation in a 
nematic liquid crystal are introduced and related to similar systems of equations in other physical contexts.  
In Section \ref{s:disp_hydro} the dispersive-hydrodynamic properties of these nematic equations are analysed 
and it is found that, while the dispersionless limit is described by a hyperbolic system equivalent to the 
shallow water equations, which is consistent with the dispersionless limit of the defocusing NLS equation, 
the linear dispersion relation is non-convex, implying the possibility of the formation of a KdV-type DSW 
in the low frequency region and the generation of high frequency resonant radiation by the DSW.  
This effect is a counterpart of the well known radiating solitary waves in systems with higher order dispersion 
studied previously in many physical contexts, from gravity-capillary waves \cite{boyd} to optical supercontinuum 
generation (see e.g.\ \cite{supercontinuum} and references therein).  In Section \ref{s:kdv} the fifth order 
KdV equation (also known as the Kawahara equation) is derived from the nematic equations under a balance between 
strong nonlocality and the small amplitude, long wave approximation.  The coefficient of the fifth order 
dispersion term is proportional to the nonlocality squared.  It is then shown numerically that the effect of 
the nonlocality on the DSW is the generation of a radiative wavetrain ahead of the DSW.  In contrast 
to the well studied radiating solitons of the fifth order KdV equation, which are intrinsically unsteady, 
the solitary wave at the leading edge of the radiating DSW remains steady due to energy influx 
from the rest of the DSW.  It can then be well approximated by the standard KdV soliton if the 
higher order dispersive term is sufficiently small.  In contrast to previous work \cite{nembore} it is found
that the velocity of the leading edge of the KdV DSW is given by a classical shock jump condition, rather than
the conservation of Riemann invariants \cite{el2}.   This suggests that the resonant wavetrain acts as
an effective viscous loss term for the DSW.  In Section 5\ref{s:wkb} a WKB solution is constructed 
for the rapidly oscillating, resonant, linear radiative wavetrain in the full nematic system under the assumption 
that the lead solitary wave in the DSW can be approximated by a KdV soliton.  Section \ref{s:num} is 
devoted to comparisons of the constructed modulation solution with full numerical solutions of the nematic 
system.  

\section{Nematic equations}
\label{s:nemeqn}

In this paper, we consider the propagation of a polarised, coherent beam of light 
through the medium of a nematic liquid crystal \cite{PR,Wiley}.
We assume that the electric field of the light is 
in the $x$ direction and that the beam propagates in the $z$ direction.  
Nematic molecules are elongated molecules, hence their name as nematic
comes from the Greek word for thread, along which electrons can move
freely.  Hence an electric field, either an external static electric field
or the electric field of light, results in the nematic molecules becoming
dipoles and rotating in the direction of the electric field due to the resulting torque 
in order to minimise the potential energy \cite{PR,Wiley}.  The molecules rotate until 
the elastic forces balance the electrostatic forces.  This rotation changes the refractive 
index of the nematic medium.  Normally a nematic is a focusing medium, so that the 
refractive index increases on rotation of the molecules.  This self-focusing can 
then balance the diffractive spreading of a light beam, so that a bright optical 
solitary wave, termed a nematicon, can form \cite{assantokhoo,PR,Wiley}.
However, the addition of azo-dyes to the nematic medium changes its structure
so that it can become a defocusing medium as rotation of the molecules then decreases 
the refractive index \cite{gaetanodark}.  In this case, a dark solitary
wave, a dark nematicon, can form, a dip in a uniform background, rather than the
rise from a background of a bright nematicon in the focusing case.  The added complication 
of the nematic medium is that if the nematic molecules are initially aligned with 
their axis, termed the director, orthogonal to the electric field, the optical 
Fre\'edericksz threshold exists so that a minimum electric field strength is required 
to overcome the elastic forces of the nematic medium before the molecules can rotate 
\cite{khoo,PR}.  To enable nematicons to form at milliwatt 
power levels so that there is not excessive heating of the nematic, which can result in
it undergoing a phase transition, an external static electric field is applied
to pre-tilt the nematic molecules at an angle $\theta_{0}$ to the $z$ direction.
In the particular case $\theta_{0} =\pi/4$, the Fre\'edericksz threshold vanishes 
\cite{assantokhoo,PR}.

Let us denote the extra rotation from the pre-tilt caused by the electric field of the 
light beam to be $\theta$.  Then in the paraxial, slowly varying envelope approximation 
the system of equations governing the propagation of a nonlinear light beam through a 
defocusing nematic liquid crystal is \cite{conti2,gaetanodark,PR,Wiley}
\begin{eqnarray}
 i \frac{\partial u}{\partial z} + \frac{1}{2}\frac{\partial^{2} u}{\partial x^{2}} - 2\theta u & = & 0 ,
\label{e:eeqn} \\
\nu \frac{\partial^{2}\theta}{\partial x^{2}} - 2q\theta & = & - 2|u|^{2} . \label{e:direqn}
\end{eqnarray}
Here $u$ is the complex valued envelope of the electric field of the light beam.
The parameter $\nu$, termed the nonlocality, measures the elastic response of the nematic and is large,
$\nu = O(100)$, in experiments \cite{waveguide}.  This large value of the nonlocality $\nu$ 
will be found to have a dominant effect on the structure of a DSW in a defocusing 
nematic liquid crystal.  The parameter $q$ is proportional to the square of the 
pre-tilting electric field.  The electric field equation (\ref{e:eeqn}) is
a nonlinear Schr\"odinger (NLS)-type equation, which is coupled to 
equation (\ref{e:direqn}) for the response of the nematic medium.

The context of the system of equations (\ref{e:eeqn}) and (\ref{e:direqn}) has been 
explained in detail in terms of the nonlinear optics of liquid crystals.  However,
this system arises in a wide range of applications.  In nonlinear optics, it
arises whenever the response of the optical medium is based on some type of diffusive
phenomenon \cite{alejandro1}, for example it arises in the optics of nonlinear 
thermal media \cite{kuz,trillo6}, for example lead glasses 
\cite{lead,segev,segev2}, and certain photorefractive crystals \cite{photo}.
A similar system of equations arises in simplified models of fluid turbulence 
\cite{alpha1} and quantum gravity \cite{grav}.  

In this paper we consider the Riemann problem for the nematic system (\ref{e:eeqn}) and (\ref{e:direqn}).
The electric field equation (\ref{e:eeqn}) will be solved with the initial condition
\begin{equation}
 u = \left\{ \begin{array}{cc}
              u_{3}, & x < 0 \\
              u_{1}, & x > 0
             \end{array}
     \right. 
\label{e:ic}
\end{equation}
at $z=0$, with $u_{3} > u_{1}$ so that a DSW is be generated.  For consistency,
the director equation (\ref{e:direqn}) gives
\begin{equation}
 \theta = \left\{ \begin{array}{cc}
              \Theta_{3} = \frac{u_{3}^{2}}{q}, & x < 0 \\
              \Theta_{1} = \frac{u_{1}^{2}}{q}, & x > 0
             \end{array}
     \right. 
\label{e:icdir}
\end{equation}
at $z=0$.

\begin{figure}
\centering
\includegraphics[width=0.33\textwidth,angle=270]{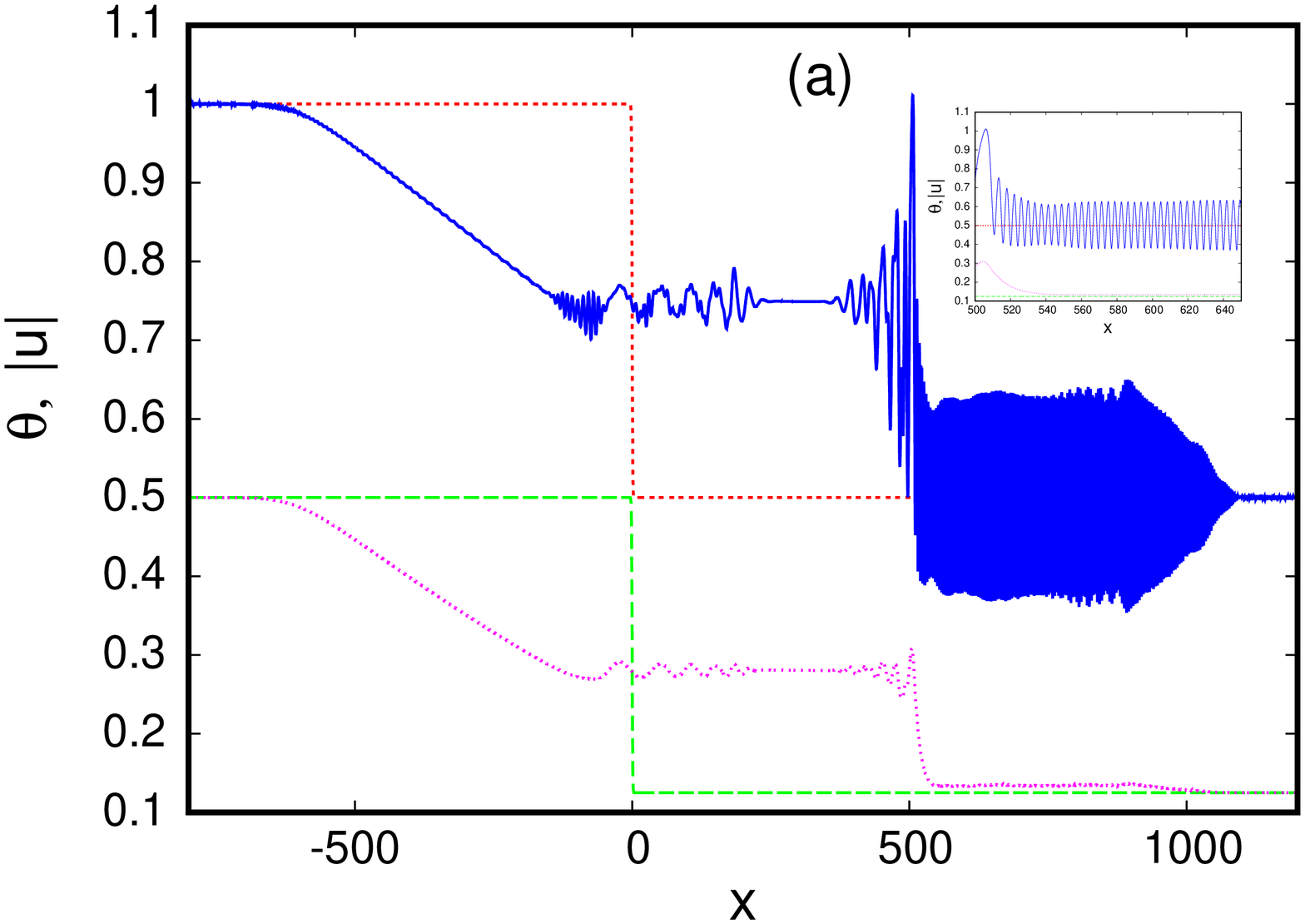}
\includegraphics[width=0.33\textwidth,angle=270]{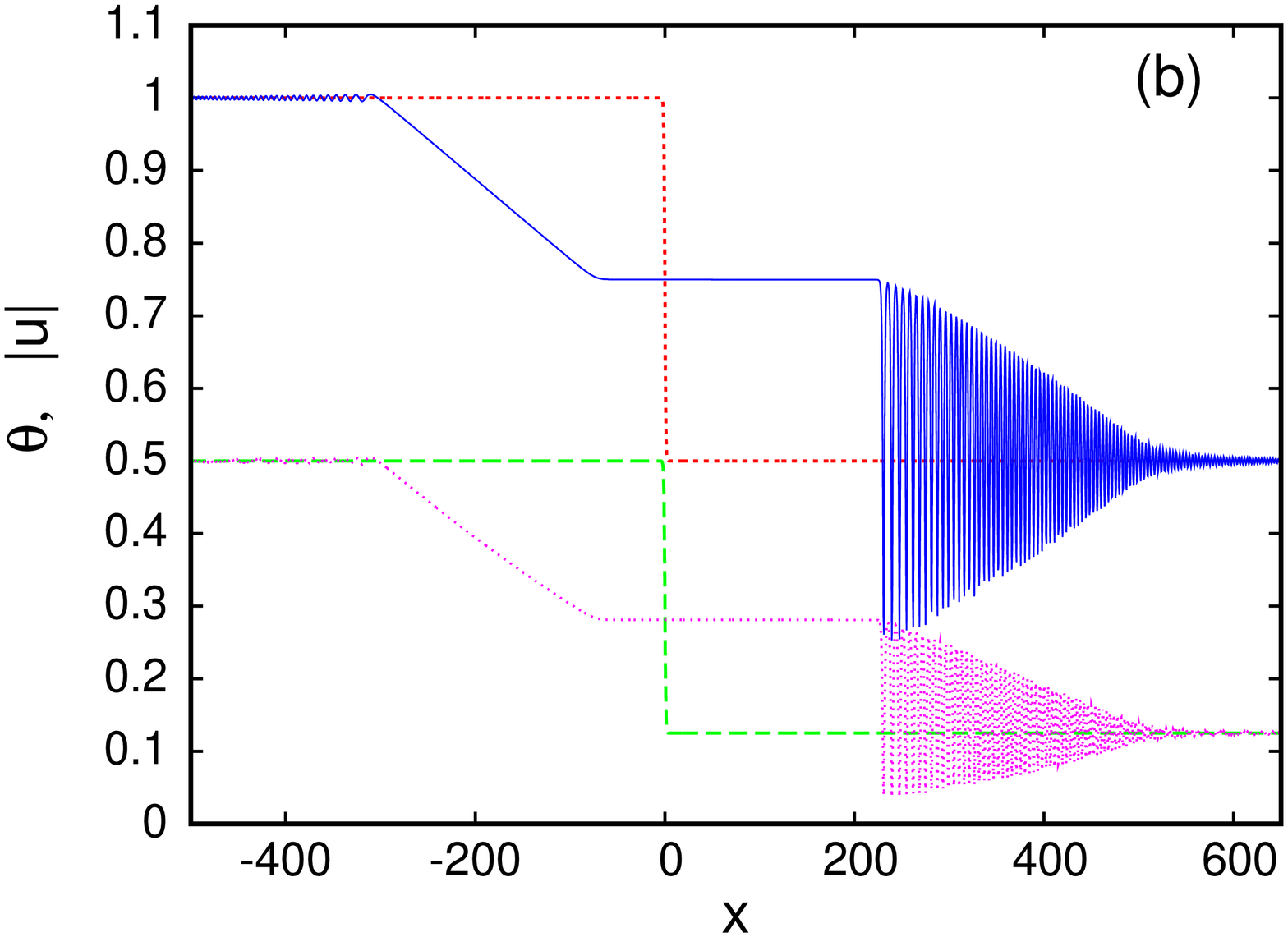}
\caption{Numerical solution of nematic equations (\ref{e:eeqn}) and (\ref{e:direqn})
for $u_{3}=1.0$, $u_{1}=0.5$ and $q=2$.  Initial condition for $|u|$ ($z=0$):  red (short dashed) line; 
initial condition for $\theta$ ($z=0$):  green (long dash) line; numerical solution for $|u|$ at $z=z_{f}$:  
blue (solid) line; numerical solution for $\theta$ at $z=z_{f}$: pink (dotted) line.  (a) $\nu=200$, 
$z_{f}=600$, inset detail of resonant wavetrain, (b) $\nu = 0.1$, $z_{f}=300$.}
\label{f:solnu1p5}
\end{figure}

A typical solution of the nematic equations for the step initial condition (\ref{e:ic}) for large 
nonlocality $\nu$ is displayed in Figure \ref{f:solnu1p5}(a).  For comparison, the solution for 
small $\nu$ is displayed in Figure \ref{f:solnu1p5}(b), noting that the nematic equations (\ref{e:eeqn}) and 
(\ref{e:direqn}) reduce to the NLS equation in the limit $\nu \to 0$.  As found in previous work \cite{nembore}, 
for large values of the nonlocality $\nu$ the solution does not display the typical defocusing NLS DSW 
structure of Figure \ref{f:solnu1p5}(b) \cite{gennady}, even though the electric field equation (\ref{e:eeqn}) 
is of defocusing NLS-type.  There is a KdV-type DSW in the electric field on the intermediate shelf of 
height $u_{2}$ between the initial levels $u_{3}$ and $u_{1}$.  Preceding this DSW, there is a relatively 
high frequency wavetrain, with a front which brings it back to the initial level $u_{1}$.  The KdV DSW and 
resonant wavetrain are mirrored in the director response, at a much reduced amplitude, with the resonant 
wavetrain in the director having amplitude $O(\nu^{-1})$.  The inset in Figure \ref{f:solnu1p5}(a) shows the
details of this resonant wavetrain.  In this paper, the complex wave structure seen 
in Figure \ref{f:solnu1p5}(a) is understood as a radiating DSW.  Such radiating DSWs typically 
arise for nonlinear wave equations with higher order dispersion, the model equation being the fifth order KdV 
equation, or Kawahara equation.  Although the theory of radiating solitons for the fifth order KdV and similar 
equations is well understood, see e.g.\ \cite{grim93,radiating_sol1,radiating_sol2} and references therein, the 
counterpart for DSW theory has only started to be explored (see the monograph \cite{bakholdin} and 
references therein and the recent papers \cite{trillo1,trillores,trilloresfour,trilloreslossbore,trilloresnature}). 
In addition to the leading resonant wavetrain, there are also radiative waves on the intermediate shelf on 
which the DSW sits.  These are most likely due to internal resonances within the DSW which is a 
modulated periodic wave with a range of phase and group speeds.

\section{Nematicon dispersive hydrodynamics}
\label{s:disp_hydro}

To analyse the Riemann problem (\ref{e:eeqn})--(\ref {e:icdir}) it is instructive to introduce the Madelung 
transformation 
\begin{equation}
 u = \sqrt{\rho} e^{i\phi}, \quad v = \phi_{x} 
\label{e:med}
\end{equation}
in order to set the nematic equations (\ref{e:eeqn}) and (\ref{e:direqn}) in the so-called dispersive 
hydrodynamic form 
\begin{eqnarray}
 \frac{\partial \rho}{\partial z} + \frac{\partial }{\partial x} \left( \rho v \right) & = & 0, \label{e:mass} \\
 \frac{\partial v}{\partial z} + v \frac{\partial v}{\partial x} + 2\frac{\partial \theta}{\partial x} 
- \frac{\partial}{\partial x} \left[ \frac{\rho_{xx}}{4\rho} - \frac{\rho_{x}^{2}}{8\rho^{2}} \right] & = & 0,
\label{e:mom} \\
\nu \frac{\partial^{2}\theta}{\partial x^{2}} - 2q\theta & = & -2\rho . \label{e:thm}
\end{eqnarray}
The above hydrodynamic form highlights the presence of two characteristic spatial scales in the system for 
large $\nu$: the long scale $O(\nu^{1/2})$ and the short scale $O(1)$, which is consistent with the two 
distinct types of oscillatory structures observed in Fig.\ \ref{f:solnu1p5}(a).  These distinct structures 
are characterised by differing typical wavelengths and different types of dispersion, which can be 
understood by analysing the linear dispersion relation for this system.

Linearising the hydrodynamic form of the nematic equations (\ref{e:mass})--(\ref{e:thm}) around the
background levels $\bar{\rho}$, $\bar{v}$ and $\bar{\theta}$ with 
\begin{equation}
 \rho = \bar{\rho} + \tilde{\rho}, \quad v = \bar{v} + \tilde{v}, \quad \theta = \frac{\bar{\rho}}{q}
+ \tilde{\theta},
\label{e:rho1}
\end{equation}
where $|\tilde{\rho}| \ll \bar{\rho}$, $|\tilde{v}| \ll |\bar{v}|$ and $|\tilde{\theta}| \ll \bar{\rho}/q$,
gives the dispersion relation for right-propagating waves \cite{nembore}
\begin{equation}
 \omega = k\bar{v} + \frac{\sqrt{\bar{\rho}k}}{\sqrt{\nu k^{2} + 2q}} \left[ 
\frac{\nu k^{2} + 2q}{4\bar{\rho}} k^{3} + 4k \right]^{1/2} .
\label{e:disp}
\end{equation}
We note that since the dispersion relation (\ref{e:disp}) is obtained not for the original system 
(\ref{e:eeqn}) and (\ref{e:direqn}), but for its dispersive-hydrodynamic representation 
(\ref{e:mass})--(\ref{e:thm}), it does not contain the frequency shift $2\Theta_1$ due to the background 
carrier wave $\sqrt{\bar {\rho}}\exp(-2i\Theta_1z)$. 

To better understand the dispersive properties of the nematic system given by the 
dispersion relation (\ref{e:disp}) we consider its long wave and short wave expansions.  
Expanding (\ref{e:disp}) in powers of $k \ll 1$ and retaining terms up to $O(k^5)$ we have
\begin{equation}
 \omega \simeq k(c +  \bar{v}) -  \frac{c}{4} \left( \frac{\nu }{q} - \frac{q}{4 \bar \rho} \right)  k^{3} 
  + \frac{c}{32}\left( \frac{3 \nu^2}{q^2} + \frac{\nu}{\bar \rho} - \frac{q^2}{16 \bar \rho^2} \right) k^5,
 \label{e:kdv5disp}
\end{equation}
where $c=\sqrt{2 \bar \rho/q}$.  The expansion (\ref{e:kdv5disp}) requires not just 
that $k \ll 1$, but that $\nu k^2 =O(1)$ or $\nu k^2 \ll 1$, which generally does not hold true even 
for reasonably small wavenumbers $k$ due to the very large value of the nonlocality 
$\nu$.  Nevertheless, as we shall see, the expansion (\ref{e:kdv5disp}) captures 
some key qualitative features of the full dispersion relation.  Now looking at the 
short wave asymptotics of (\ref{e:disp}), we obtain that for strong nonlocality 
$\nu \gg 1$
\begin{equation}
\omega \simeq k \bar v + \frac12 k^2 + O((k\sqrt{\nu})^{-1}), \quad \nu k^{2} \gg 1.
\label{e:shortw}
\end{equation}
One can see from the expansions (\ref{e:kdv5disp}) and (\ref{e:shortw}) that for sufficiently small wavenumbers 
$\omega_{kk}<0$, while for large wavenumbers $\omega_{kk}>0$.  Thus the full dispersion relation (\ref{e:disp}) 
is non-convex, which has important physical consequences as it implies the possibility of resonance between long 
and short waves and hence the generation of short wave radiation propagating ahead of the DSW.  The 
effect of resonant radiation generation by solitary waves in equations with higher order dispersion is well 
known in the context of gravity-capillary waves, see e.g.\ \cite{boyd, grim93, radiating_sol1,radiating_sol2} 
and references therein.  There is also abundant literature on radiating solitary waves in nonlinear optics, 
see e.g.\ \cite{kivshar, supercontinuum} and references therein.  However, the counterpart of this for 
DSW theory is yet to be developed.  A few existing notable contributions include numerical investigations of 
radiating DSWs described in the monograph \cite{bakholdin} and the recent papers 
\cite{trillo1,trillores,trilloresfour,trilloreslossbore,trilloresnature} on the effects of higher order 
dispersion on NLS DSWs in the context of nonlinear optics.

\begin{figure}
\centering
\includegraphics[width=0.33\textwidth,angle=270]{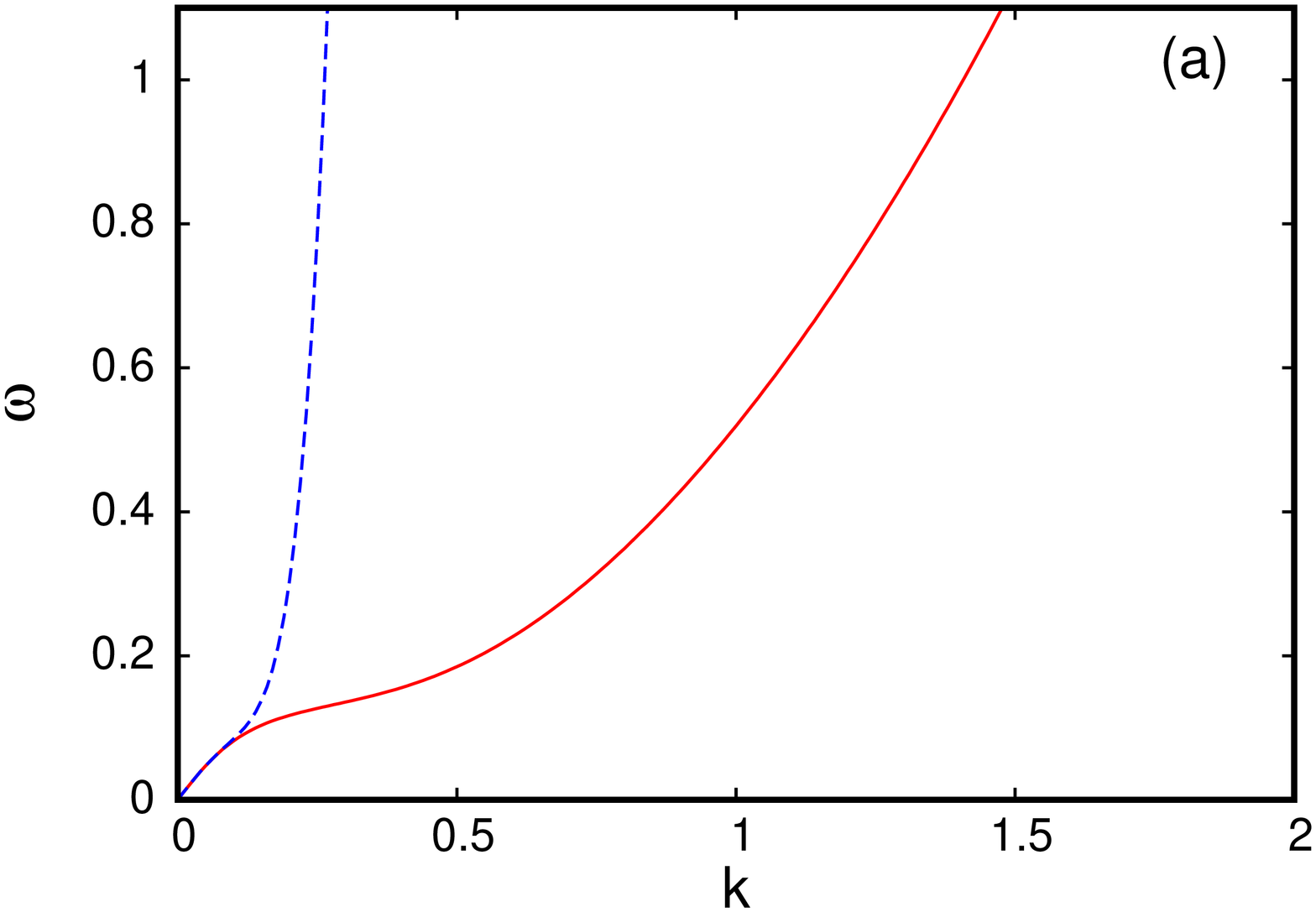}
\includegraphics[width=0.33\textwidth,angle=270]{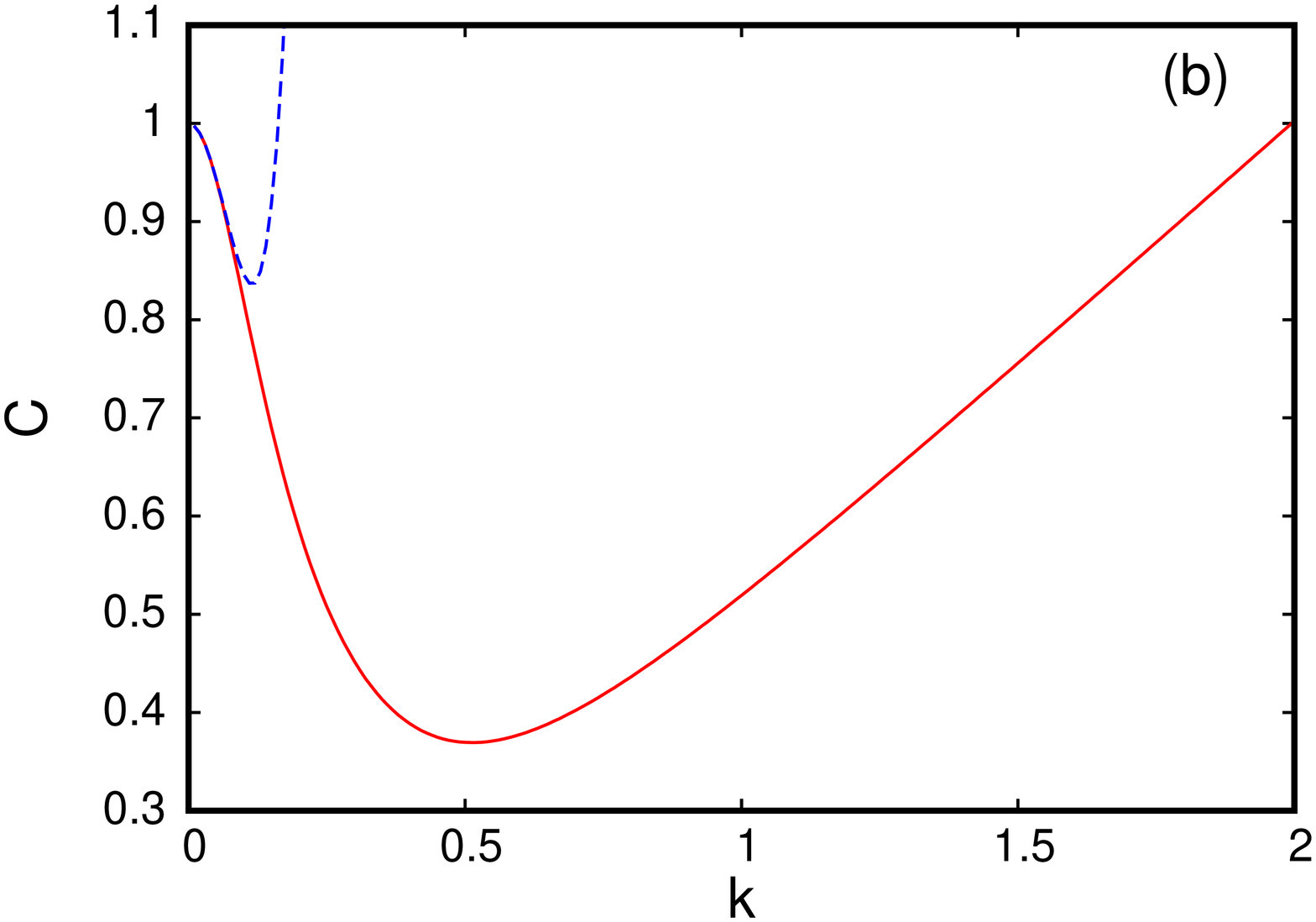}
\caption{Nematic dispersion relation (a) Full dispersion relation (\ref{e:disp}):  
red (solid) line, 5th order expansion (\ref{e:kdv5disp}):  blue (dashed) line, (b)
nematic phase velocity $C = \omega / k$:  red (solid) line, 4th order expansion 
velocity:  blue (dashed) line.  The parameters are $\bar{\rho}=1.0$, $\bar{v}=0$, 
$\nu = 200$ and $q=2$.}
\label{f:kdv5}
\end{figure}

In Figure \ref{f:kdv5} a comparison between the full dispersion relation (\ref{e:disp}) and the 5-th order 
Taylor expansion (\ref{e:kdv5disp}) is shown for the physically realistic nonlocality $\nu =200$ \cite{waveguide}.   
It can be seen that (\ref{e:kdv5disp}) is a good approximation to the full dispersion relation in the limit of 
low wavenumber, as expected.  However, due to the large factor in front of the $k^{5}$ term in the approximate 
dispersion relation (\ref{e:kdv5disp}) the low wavenumber expansion rapidly deviates from the exact dispersion 
relation as $k$ increases.  Nevertheless, it qualitatively captures the key feature of the full dispersion relation, 
its non-convexity, so can be used for qualitative predictions of the effects of nonlocality on the nematic 
DSW behaviour.  It is further seen from Figure \ref{f:kdv5} that the full phase velocity $\omega/k$ is not monotone 
and has a minimum, which is also qualitatively captured by the long wave dispersion relation (\ref{e:kdv5disp}).  
The corresponding nonlinear equation with this linear dispersion relation, the fifth order KdV equation, will be 
derived in the next section.

Let us now look at the opposite, dispersionless limit of the nematic system (\ref{e:mass})--(\ref{e:thm}), which is 
described by the hyperbolic system of shallow water type \cite{whitham}
\begin{eqnarray}
 \frac{\partial \rho}{\partial z} + \frac{\partial }{\partial x} \left( \rho v \right) & = & 0, \label{e:massnd} \\
 \frac{\partial v}{\partial z} + v \frac{\partial v}{\partial x} + 2\frac{\partial \theta}{\partial x} 
& = & 0, \label{e:momnd} \\
 \theta & = & \frac{\rho}{q} . \label{e:thmnd} 
\end{eqnarray}
These equations can be set in the Riemann invariant form
\begin{eqnarray}
 & & v + \frac{2\sqrt{2}}{\sqrt{q}} \sqrt{\rho} = \mbox{constant} 
 \quad \mbox{on} \quad C_{+}:  \frac{dx}{dz} = V_{+} = v + \frac{\sqrt{2}}{\sqrt{q}} \sqrt{\rho},
\label{e:cp} \\
& & v - \frac{2\sqrt{2}}{\sqrt{q}} \sqrt{\rho}= \mbox{constant} 
\quad \mbox{on} \quad C_{-}:  \frac{dx}{dz} = s_{+} = v - \frac{\sqrt{2}}{\sqrt{q}} \sqrt{\rho} .
\label{e:cm} 
\end{eqnarray}
The rarefaction wave seen in Fig.\ \ref{f:solnu1p5} can then be described by a centred 
simple wave solution of equations (\ref{e:cp}) and (\ref{e:cm}) in which the right-going 
Riemann invariant is constant.  This solution will be presented in Section 
5\ref{s:nondisp}.

\section{Fifth order KdV equation}
\label{s:kdv}

It has been shown that the nematic system (\ref{e:eeqn}) and (\ref{e:direqn})
reduces to the Korteweg-de Vries (KdV) equation in the limit of small deviations from a 
background level \cite{hor,nembore}.  However, the physically large value of the nonlocality 
$\nu$ \cite{waveguide} and the linked resonant wavetrain have major effects on the asymptotic analysis, 
which were not considered in this previous work.

Indeed, assuming $\nu \gg 1$, but $\nu k^2 = O(1)$, one has to retain the fifth order terms 
in the dispersion relation expansion (\ref{e:kdv5disp}), which implies the necessity of 
keeping the fifth order dispersion term in the corresponding asymptotic KdV equation.  
The asymptotic reduction of the nematic equations to the KdV equation in the limit of 
small deviations from a background level $u_{0}$ will then be revisited, taking account 
of the large value of the nonlocality $\nu$.  This asymptotic KdV equation will be 
derived from the hydrodynamic form of the nematic equations (\ref{e:mass})--(\ref{e:thm}).  
Let us expand the hydrodynamic variables as 
\begin{eqnarray}
 \rho & = & \rho_{0} + \epsilon^{2} P_{1}(\xi,\eta) + \epsilon^{4}P_{2}(\xi,\eta) + \ldots, \label{e:ukdv} \\
 v & = & \epsilon^{2}V_{1}(\xi,\eta) + \epsilon^{4} V_{2}(\xi,\eta)
+ \epsilon^{6}V_{3}(\xi,\eta) + \ldots, \label{e:phikdv} \\
 \theta & = & \frac{\rho_{0}}{q} + \epsilon^{2}\theta_{1}(\xi,\eta) + \epsilon^{4}\theta_{2}(\xi,\eta) 
+ \epsilon^{6}\theta_{3}(\xi,\eta) + \ldots, \label{e:thetakdv}
\end{eqnarray}
where $\rho_{0} = u_{0}^{2}$ and $0 < \epsilon \ll 1$ is a measure of the deviation from the background 
$u_{0}$, here $\epsilon^{2} = u_{2}-u_{1}$.  Also, $\xi = \epsilon \left( x - U z \right)$ and 
$\eta = \epsilon^{3} z$
are the usual stretched variables used to derive the KdV equation \cite{whitham}.  We also 
assume that all corrections to the equilibrium state $\rho= \rho_0$, $v=0$, 
$\theta=\rho_0/q$ vanish as $|\xi| \to \infty$.

Substituting the expansions (\ref{e:ukdv}) and (\ref{e:thetakdv}) into the 
director equation (\ref{e:direqn}), we obtain at $O(\epsilon^{2})$ 
\begin{equation}
 \theta_{1} = \frac{P_{1}}{q}
 \label{e:thep2}
\end{equation}
and at $O(\epsilon^{4})$
\begin{equation}
\theta_{2} =  \frac{\nu}{2q} \frac{\partial^{2}\theta_{1}}{\partial \xi^{2}}
+ \frac{P_{2}}{q}  + \frac{\nu \epsilon^{2}}{2q} \frac{\partial^{2} \theta_{2}}
{\partial \xi^{2}} .
 \label{e:thep4}
\end{equation}
The term $\nu \epsilon^{2}\theta_{2\xi\xi}/2q$ is formally $O(\epsilon^{2})$ and
should appear at next order in the expression for $\theta_{3}$, as in \cite{hor}.  However, 
this implicitly assumes that $\nu = O(1)$, which is not the case for experimental values of $\nu$.  
Hence, this term will be retained at $O(\epsilon^{4})$.   Treating $\nu \epsilon^{2}\theta_{2\xi\xi}/2q$
as a correction, equation (\ref{e:thep4}) can be solved for $\theta_{2}$ to give
\begin{equation}
 \theta_{2} =  \left [ \frac{\nu}{2q} \frac{\partial^{2}\theta_{1}}{\partial \xi^{2}}
+ \frac{P_{2}}{q}  \right ] +  \frac{\nu^{2} \epsilon^{2}}{4q^{2}} \frac{\partial^{4} \theta_{1}}
{\partial \xi^{4}} +    \frac{\nu \epsilon^2}{2q^2} \frac{\partial ^2 P_2}{\partial \xi^2} .
 \label{e:thep4r2}
\end{equation}
Note that the last term in (\ref{e:thep4r2}) has to be retained as (\ref{e:thep4}) 
implies that $P_2$ can be of $O(\nu)$, making the last term $O(\nu^2 \epsilon^2)$. 

Substituting the expansions (\ref{e:ukdv})--(\ref{e:thetakdv}) into the
``mass'' and ``momentum'' equations (\ref{e:mass}) and (\ref{e:mom}), we have at $O(\epsilon^{3})$
\begin{equation}
 \frac{\partial V_{1}}{\partial \xi} = \frac{U}{\rho_{0}} \frac{\partial P_{1}}{\partial \xi}
\quad \mbox{and} \quad U \frac{\partial V_{1}}{\partial \xi} = \frac{2}{q} \frac{\partial P_{1}}{\partial \xi},
\label{e:phi1}
\end{equation}
respectively, on using (\ref{e:thep2}) for $\theta_{1}$.  Compatibility between these two equations
for $V_{1}$ and $P_{1}$ then gives the coordinate velocity $U$ as
\begin{equation}
U^{2} = \frac{2}{q} \rho_{0}.
 \label{e:cexp}
\end{equation}
Identifying $u_0^2 = \bar \rho$ from Section \ref{s:disp_hydro}, we see that $U=c$ from the 
long wave expansion (\ref{e:kdv5disp}) of the linear dispersion relation.

Similarly, at $O(\epsilon^{5})$ the mass and momentum equations (\ref{e:mass}) and (\ref{e:mom})
give
\begin{equation}
\rho_{0} \frac{\partial V_{2}}{\partial \xi}
 -U \frac{\partial P_{2}}{\partial \xi} + \frac{\partial P_{1}}{\partial \eta}
 + V_{1}\frac{\partial P_{1}}{\partial \xi} + P_{1} \frac{\partial V_{1}}{\partial \xi} = 0 
 \label{e:ep5}
\end{equation}
and
\begin{equation}
-U \frac{\partial V_{2}}{\partial \xi} + 2\frac{\partial \theta_{2}}{\partial \xi} + 
\frac{\partial V_{1}}{\partial \eta} + V_{1} \frac{\partial V_{1}}{\partial \xi} 
- \frac{1}{4\rho_{0}} \frac{\partial^{3} P_{1}}{\partial \xi^{3}} = 0, 
\label{e:ep4}
\end{equation}
respectively.  

It was shown in \cite{nembore,hor} that substituting the leading order part of 
(\ref{e:thep4r2}) (the terms in brackets) into (\ref{e:ep4}) and combining it with 
(\ref{e:phi1}) and (\ref{e:ep5}) leads to the KdV equation.  We now need to extend 
this derivation by including the higher order terms of (\ref{e:thep4r2}).  The problem 
we encounter is with the computation of the last term in (\ref{e:thep4r2}) as the 
correction $P_2$ cannot be computed separately at order $O(\epsilon^5)$, leading 
to equations (\ref{e:ep5}) and (\ref{e:ep4}), and a higher order approximation is required. 
This difficulty can be circumvented by suggesting a suitable ${\it ansatz}$ for $P_2$ 
which is compatible with (\ref{e:ep5}) and (\ref{e:ep4}).  Let
\begin{equation}\label{e:p2}
P_2=\alpha\nu \frac{\partial^2 \theta_1}{\partial \xi^2} =  \alpha \frac{ \nu}{q} 
\frac{\partial^2 P_1}{\partial \xi^2},
\end{equation}
where $\alpha$ is a constant. Then substituting (\ref{e:thep4r2}) and (\ref{e:p2})  
into (\ref{e:ep4}) we obtain, on using (\ref{e:phi1}),
\begin{equation} \label{e:v2}
\frac{\partial V_2}{\partial \xi} = - \frac{1}{\rho_0} \left[  \frac{\partial P_1}{\partial \eta} + 
\frac{2U}{\rho_0} P_1 \frac{\partial P_1}{\partial \xi}  -  \frac{\alpha  \nu U}{q} 
\frac{\partial^3 P_1}{\partial \xi^3} \right] .
\end{equation}

Substituting (\ref{e:thep4r2}), (\ref{e:p2}) and (\ref{e:v2}) into (\ref{e:ep5}) 
we obtain the fifth order KdV equation for $P_{1}$
\begin{equation}
\frac{\partial P_{1}}{\partial \eta} + \frac{3}{qU} P_{1}\frac{\partial P_{1}}{\partial \xi}
+ \frac{U}{4}\left( \frac{\nu }{q} - \frac{q}{4\rho_{0}} \right) \frac{\partial^{3}P_{1}}{\partial \xi^{3}} 
+ \frac{\nu^{2}\epsilon^{2}\rho_{0}}{4q^{3}U} (1+\alpha) \frac{\partial^{5} P_{1}}{\partial \xi^{5}} = 0.
\label{e:kdv5}
\end{equation}
For the 5th order KdV equation (\ref{e:kdv5}) to be consistent with the long wave 
expansion (\ref{e:kdv5disp}) of the linear dispersion relation \cite{whitham} we have 
to choose $\alpha=-1/4$ (note that due to the scaling for $\xi$ and $\eta$ one has to replace 
$(\omega - kc) \to \epsilon^3 \omega $, $k \to \epsilon k$ in (\ref{e:kdv5disp}) to 
make the comparison).  We note that if the substitution (\ref{e:p2}) were not compatible 
with equations (\ref{e:ep5}) and (\ref{e:ep4}), it would not be possible to obtain agreement 
for both dispersive terms in (\ref{e:kdv5}) with the expansion of the nematic 
dispersion relation (\ref{e:kdv5disp}) using the single fitting parameter $\alpha$.

The 5th order KdV equation (\ref{e:kdv5}) differs from that found in \cite{hor,nembore} due to the 
$P_{1\xi\xi\xi\xi\xi}$ term, which arises at this order as $\nu$ is large.  The polarity of the solitary wave 
solution of the 5th order KdV equation (\ref{e:kdv5}) depends on the sign of the coefficient of the 
$P_{1\xi\xi\xi}$ term.  It is then clear that in the nonlocal regime with $\nu$ large the solitary wave 
solution of the defocusing nematic equations (\ref{e:eeqn}) and (\ref{e:direqn}) is a 
bright solitary wave, rising above a background level, rather than the usual dark solitary 
wave of the defocusing NLS equation, which the nematic equations become in the limit 
$\nu \to 0$.  

Although the fifth order KdV equation (\ref{e:kdv5}) has a limited range of validity as 
an asymptotic, quantitative model for nematic DSWs, it provides major qualitative 
insight into their dynamics by capturing the effect of resonant radiation.  To illustrate 
this, we solved numerically the normalised 5th order KdV equation
\begin{equation}
 \frac{\partial w}{\partial t} + 6w\frac{\partial w}{\partial x}
+ \frac{\partial^{3}w}{\partial x^{3}} + \gamma \frac{\partial^{5} w}{\partial x^{5}} = 0
\label{e:kdv5norm}
\end{equation}
for sufficiently small $\gamma>0$.  Equation (\ref{e:kdv5norm}) has been derived in several physical contexts, 
including magnetoacoustic waves and capillary-gravity waves of small amplitude when the Bond number is close to, 
but just less than, $1/3$ (see e.g. \cite{boyd} and references therein).  Radiating solitary waves solutions of 
(\ref{e:kdv5norm}) were discovered by Kawahara \cite{kawahara} and then studied analytically 
and numerically in a number of papers (see e.g.\ \cite{grim93, radiating_sol1, radiating_sol2} and references 
therein).

\begin{figure}
\centering
\includegraphics[width=0.5\textwidth,angle=270]{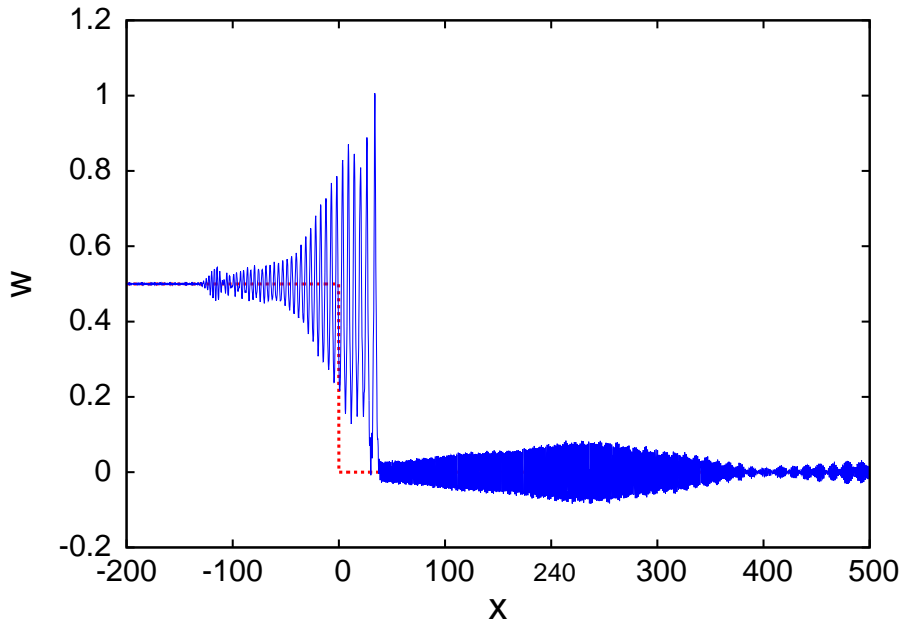}
\caption{Numerical solution of 5th order KdV equation (\ref{e:kdv5norm}) for $w_{0}=0.5$ and
$\gamma=0.05$.  red (dashed) line:  initial condition at $t=0$, blue (solid) line:  
solution at $t=20$.}
\label{f:kdv5soln}
\end{figure}

Let us consider the 5th order KdV equation (\ref{e:kdv5norm}) with the initial condition $w = 0$, $x>0$ and 
$w=w_{0}$, $ x < 0$, so that a DSW is generated.  Due to the non-convexity of the dispersion relation 
for (\ref{e:kdv5norm}), there is the possibility of energy exchange between long and short waves propagating 
with the same phase velocity (see Figure \ref{f:kdv5}), so this DSW is expected to generate a resonant linear 
wavetrain propagating ahead of it \cite{bakholdin}.  Such a radiating KdV DSW is displayed in Figure 
\ref{f:kdv5soln}.  The solution shown in this figure has strong similarities to the 
radiating nematic DSW solution of Figure \ref{f:solnu1p5}(a).  However, the resonant 
wavetrain of the nematic solution is more uniform, which is due to the smoothing 
effect of the large nonlocality $\nu$.

In conclusion, we note that, although the known theory of radiating solitary waves provides some intuition as to the 
counterpart radiating DSW solution, the major contrasting feature of radiating DSWs is the 
fact that the lead solitary wave of the radiating DSW remains steady, while an isolated radiating solitary wave 
is intrinsically unsteady due to the radiation carrying away the solitary wave's energy \cite{grim93}.  

\section{Dam break problem for the nematic system}
\label{s:riemann}

The solution of the Riemann problem (\ref{e:eeqn})--(\ref{e:icdir}) for the nematic system generically consists 
of three distinct parts: a rarefaction wave, a (bright) DSW and a radiative wavetrain (see Figure 
\ref{f:solnu1p5}(a)).  The rarefaction wave was analysed in \cite{nembore}, so below we only briefly outline the 
relevant results.  Our main attention in this section will be on the DSW on the intermediate 
level $u_{2}$ and the resonant wavetrain generated by it.
 
\subsection{Rarefaction wave}
\label{s:nondisp}

The solution displayed in Figure \ref{f:solnu1p5}(a) shows that there is an expansion wave
linking the initial level $u_{3}$ behind the DSW to the level $u_{2}$ on which the KdV 
DSW sits.  This expansion wave solution has already been determined \cite{nembore}, so 
only the relevant details will be given here.  

The expansion wave linking the initial level $u_{3}$ behind to the intermediate level 
$u_{2} = \sqrt{\rho_{2}}$ can be found as a simple wave solution of the Riemann invariant 
equations (\ref{e:cp}) and (\ref{e:cm}) as \cite{nembore}
\begin{equation}
\sqrt{\rho} = \left\{ \begin{array}{cc}
                        u_{3}, & \frac{x}{z} < -\frac{\sqrt{2}u_{3}}{\sqrt{q}} \\
                        \frac{\sqrt{q}}{3\sqrt{2}} \left[ \frac{2\sqrt{2}u_{3}}{\sqrt{q}} - \frac{x}{z} \right], &
-\frac{\sqrt{2}u_{3}}{\sqrt{q}} \le \frac{x}{z} \le \frac{\sqrt{2}}{\sqrt{q}} \left( 2u_{3} - 3\sqrt{\rho_{2}}
\right) \\
\sqrt{\rho_{2}}, & \frac{\sqrt{2}}{\sqrt{q}} \left( 2u_{3} - 3\sqrt{\rho_{2}} \right) < \frac{x}{z} \le s_{+}
\end{array}
\right.
\label{e:midsolnu}
\end{equation}
and
\begin{equation}
v = \left\{ \begin{array}{cc}
                        0, & \frac{x}{z} < -\frac{\sqrt{2}u_{3}}{\sqrt{q}} \\
                        \frac{2\sqrt{2}u_{3}}{3\sqrt{q}} + \frac{2x}{3z}, &
-\frac{\sqrt{2}u_{3}}{\sqrt{q}} \le \frac{x}{z} \le \frac{\sqrt{2}}{\sqrt{q}} \left( 2u_{3} - 3\sqrt{\rho_{2}}
\right) \\
v_{2} = \frac{2\sqrt{2}}{\sqrt{q}} \left( u_{3} - \sqrt{\rho_{2}} \right), & 
\frac{\sqrt{2}}{\sqrt{q}} \left( 2u_{3} - 3\sqrt{\rho_{2}} \right) < \frac{x}{z} \le s_{+}
\end{array}
\right. .
\label{e:midsolnv}
\end{equation}
Here $s_{+}$ is the velocity of the lead soliton of the KdV DSW, which lies at the leading edge of 
the intermediate shelf.  The simple wave solution (\ref{e:midsolnu}) linking the initial level $u_{3}$ 
and the intermediate shelf $u_{2}$ will be used for the comparisons with numerical solutions 
in Section \ref{s:num}.

The velocity $s_{+}$ of the leading edge of the DSW on the shelf $u_{2}$ needs to
be determined.  In contrast to the KdV and defocusing NLS equations, this velocity is not
determined by the conservation of the Riemann invariant on $C_{-}$ across the DSW \cite{el2},
but by the classical shock jump condition.  We note here that the occurrence of the classical 
shock conditions in a conservative dispersive hydrodynamics was observed earlier in 
numerical simulations of large amplitude shallow water DSWs \cite{gn} and 
optical DSWs in photorefractive media \cite{photoref} (see 
also \cite{hoefer_euler}) and, very recently, in the context of radiating dispersive 
shock waves governed by the defocusing NLS equation modified by third order dispersion 
\cite{trillo1,trillores,trilloresfour,trilloresnature}.  This remarkable generic phenomenon 
requires further analytical study.  

The non-dispersive equations (\ref{e:massnd})--(\ref{e:thmnd}) have the jump conditions \cite{whitham}
\begin{equation}
 s_{+} = \frac{\rho_{2} v_{2}}{\rho_{2} - \rho_{1}} \quad \mbox{and} \quad s_{+} = \frac{\frac{1}{2}v_{2}^{2}
 + \frac{2\rho_{2}}{q} - \frac{2\rho_{1}}{q}}{v_{2}} 
 \label{e:jump}
\end{equation}
as ahead of the shock, $\rho = \rho_{1}$ and $v=0$ and behind the shock, $\rho = \rho_{2}$ and
$v=v_{2}$.  Eliminating between these equations gives
\begin{equation}
 v_{2} = \frac{2}{\sqrt{q}} \frac{\rho_{2}-\rho_{1}}{\sqrt{\rho_{2} + \rho_{1}}}  \quad \mbox{and} \quad
 s_{+} = \frac{2}{\sqrt{q}} \frac{\rho_{2}}{\sqrt{\rho_{2} + \rho_{1}}} .
 \label{e:v2sp}
\end{equation}
The expansion fan solution (\ref{e:midsolnv}) also gives an expression for $v_{2}$ in terms
of the intermediate level $u_{2}$.  Matching this and (\ref{e:v2sp}) then gives that this intermediate
level $u_{2} = \rho_{2}^{2}$ is the solution of
\begin{equation}
 u_{2}^{4} - 4u_{3}u_{2}^{3} + 2\left( u_{3}^{2} + 2u_{1}^{2}\right) u_{2}^{2} - 4u_{3}u_{1}^{2}u_{2}
 - u_{1}^{4} + 2u_{3}^{2}u_{1}^{2} = 0 
 \label{e:u2expr}
\end{equation}
with $u_{1} \le u_{2} \le u_{3}$.  For the particular case $u_{1}=0$, $u_{2} = (2 - \sqrt{2})u_{3}$.  
Also, as $u_{1} \to u_{3}$, $u_{2} \to \left(u_{3} + u_{1}\right)/2$, which is the value obtained by 
conservation of the Riemann invariant (\ref{e:cm}) on $C_{-}$ \cite{nembore}.

\subsection{Dispersive shock wave: lead solitary wave}
\label{s:kdvbore}

In previous work \cite{nembore} the DSW solution of the KdV equation 
\cite{gur,bengt} was used for the DSW on the intermediate shelf $u_{2}$.  While this 
was found to give good agreement with numerical solutions for values of $u_{1}$ near 
$u_{3}$, significant disagreement was found for values of $u_{1}$ away from $u_{3}$.  
As discussed above, this is due to the velocity $s_+$ of the front of the full nematic DSW  not 
being well determined by the velocity of the leading edge of the asymptotic KdV DSW.  The reason for this 
behaviour is that the DSW is subject to radiative losses due to the resonance with the co-propagating 
linear short wavelength waves, resulting in a rapidly oscillating wavetrain shed ahead of the DSW.  
For small initial steps the radiating DSW is described in the framework of the 5th order KdV 
equation (\ref{e:kdv5}).  However, for general jumps the full nematic system should be used due 
to the 5th order KdV equation not being accurate in capturing large wavenumber dispersive 
behaviour (see Fig.~\ref{f:kdv5}).

We now need to relate the shock velocity $s_{+}$ to the amplitude of the lead soliton 
of the DSW.  Since the solitary wave solution of the full nematic system is not 
available, as an approximation we shall use the soliton solution of the standard KdV 
equation, that is (\ref{e:kdv5}) without the 5th derivative term.  On noting the scalings
in the expansions (\ref{e:ukdv})--(\ref{e:thetakdv}) and equating $s_{+}$ given by
(\ref{e:v2sp}) to the lead soliton velocity, this gives
\begin{equation}
 a_s=\epsilon^{2} A = \frac{\sqrt{2} u_{2}^{2}}{\sqrt{u_{2}^{2} + u_{1}^{2}}} - u_{1} .
 \label{e:kdvboreamp}
\end{equation}
The lead soliton of the KdV DSW itself is given by \cite{nembore}
\begin{equation}
 |u| = \sqrt{\rho} = u_{0} + \epsilon^{2} A \hbox{sech}^{2} \beta (x-s_{+}z) 
 + \ldots,
\label{e:kdvlead}
\end{equation}
where
\begin{equation}
 \beta = \frac{\epsilon \sqrt{A}}{\sqrt{2}(2q)^{1/4} \alpha \sqrt{\nu}} 
 \quad \mbox{and} \quad \alpha = \left[ \frac{U}{4q} - 
 \frac{qU}{16u_{1}^{2}\nu} \right]^{1/2} .
 \label{e:beta}
\end{equation}
These results will be used in the next section to find a solution for the resonant
wavetrain leading the KdV DSW seen in Figure \ref{f:solnu1p5}(a) on identifying 
$u_{0} = u_{1}$.

\subsection{Resonant wavetrain}
\label{s:wkb}

Let us now consider the wavetrain ahead of the DSW.  This wavetrain
is generated due to a resonance between the long wave oscillations in the
DSW and co-propagating short wavelength waves, as implied by the non-convexity of the 
linear dispersion relation (\ref{e:disp}) and discussed in Section \ref{s:kdv}.

In determining the structure of the resonant wavetrain we refer to Figure \ref{f:solnu1p5}(a) 
in which one can observe three regions of distinctly different behaviour:  {\it region (i)} 
provides a transition from the lead soliton of the DSW to {\it region (ii)} which contains 
the (almost) uniform extended middle part of the wavetrain; and the front {\it region (iii)} which brings 
the wavetrain down to the constant level $u=u_1$ and $\theta = \Theta_1$, where $\Theta_1= |u_{1}|^{2}/q$, 
see (\ref{e:icdir}).

We start with the middle region (ii) for which the director $\theta$ is close to $\Theta_1$, 
$\theta - \Theta_1 = O(\nu^{-1})$ and the wavenumber $k$ is $O(1)$.  Hence, the asymptotic 
dispersion relation (\ref{e:shortw}) applies and the dispersion relation for the resonant wavetrain 
is 
\begin{equation}
 \omega_{r} = \frac{1}{2} k^{2} + 2\Theta_{1}
\label{e:dispwave}
\end{equation}
as the resonant wavetrain is on the background carrier wave $u_{1}\exp(-2i\Theta_{1}z)$.
Furthermore, the resonant wavetrain in the region (ii) is then asymptotically 
described by the linear equation following from (\ref{e:eeqn}) on setting $\theta=\Theta_1$,
\begin{equation}
 i \frac{\partial u}{\partial z} + \frac{1}{2} \frac{\partial^{2}u}{\partial x^{2}} - 2\Theta_{1} u = 0 .
\label{e:eeqnwave}
\end{equation}

We assume that the main resonance is with the lead soliton of the DSW, which we approximate 
by the KdV soliton (\ref{e:kdvlead}).  Matching the phase velocity to the lead KdV soliton velocity 
(\ref{e:v2sp}), we have 
\begin{equation}
 c_{r} = \frac{1}{2} k + \frac{2\Theta_{1}}{k} = s_{+},
\label{e:phasegroup}
\end{equation}
which can be solved to give the wavenumber of the resonant wavetrain as
\begin{equation}
 k = k_{r} = s_{+} + \left[ s_{+}^{2} - \frac{4}{q} u_{1}^{2} \right]^{1/2}.
 \label{e:kres}
\end{equation}
The front of the resonant wavetrain moves at the group velocity $c_{g}$ \cite{whitham}, 
which is
\begin{equation}
 c_{g} = \omega_r'(k_r)=k_{r}.
 \label{e:group}
\end{equation}
These expressions for the asymptotic wavenumber of the resonant wavetrain away from the 
DSW and the velocity of the front of the wavetrain will be used in the solution for this
wavetrain.

The wavenumber (\ref{e:kres}) is real if $u_{1} \le u_{1c}$, where $u_{1c}$ is the solution of
\begin{equation}
 s_{+} = \frac{2}{\sqrt{q}} \frac{u_{2}^{2}}{\sqrt{u_{2}^{2} + u_{1c}^{2}}} = \frac{2}{\sqrt{q}} u_{1c} .
 \label{e:resbound}
\end{equation}
For $u_{1}$ above $u_{1c}$ there is only a transient wavetrain ahead of the DSW 
\cite{whitham}.  This existence of a critical $u_{1}$ above which there is no resonant 
wavetrain is in agreement with previous work \cite{nembore} in which the critical value 
was found to be $u_{1c} = u_{3}/\sqrt{2}$.  For $u_{3} = 1$, $q=2$ and $\nu = 200$ 
numerical solutions give the critical value $u_{1c} = 0.69$ \cite{nembore}.  For these 
parameter values the new modulation value (\ref{e:resbound}) $u_{1c} = 0.648$ is 
slightly below the numerical cut-off, while the previous modulation value 
$u_{1c} = 1/\sqrt{2}$ is slightly above.  It should be noted that numerical solutions 
do not show a sharp transition to no resonant wavetrain as given by (\ref{e:resbound}), 
but a rapid transition from an upstream uniform wavetrain to none over a $u_{1}$ range 
of about $0.1$.  

Above the critical value (\ref{e:resbound}) the resonant wavetrain ceases to exist.  The DSW on the 
intermediate level $u_{2}$ then becomes the standard KdV type DSW and the approximate solution of 
\cite{nembore} holds.  The amplitude and velocity of the lead soliton of the DSW are, then cf.\ 
(\ref{e:v2sp}), (\ref{e:kdvboreamp}), 
\begin{equation}
 a_{s} = \epsilon^{2} A = u_{3} - u_{1} , \quad s_{+} = \sqrt{\frac{2}{q}} u_{3} . 
 \label{e:kdvboreampnores}
\end{equation}

As equation (\ref{e:eeqnwave}) is linear, it does not allow the determination of the 
resonant wavetrain amplitude.  For that, one needs to go beyond the approximation $\theta=\Theta_1$ in 
the wavetrain and include the (significant) variations of the director in the transition region (i) between 
the DSW and the uniform wavetrain region.

As the phase velocity of the resonant wavetrain is the same as the (classical shock) 
velocity $s_{+}$ (\ref{e:v2sp}) of the lead soliton of the DSW, to determine the 
solution for the wavetrain in the transition region we will use the moving coordinate $\zeta = x - s_{+}z$.  
By inspection of the numerical solution of Figure \ref{f:solnu1p5}(a) it is reasonable to assume that  
the approximate director solution in the transition region is given by the lead solitary wave 
of the DSW, so that from equations (\ref{e:thep2}) and (\ref{e:kdvlead}) of the KdV 
expansion of Sections \ref{s:kdv} and 5\ref{s:kdvbore} we have
\begin{equation}
\theta  =  \frac{u_{1}^{2}}{q} + \epsilon^{2} \frac{2u_{1}}{q}A
 \hbox{sech}^{2} \beta \zeta.
  \label{e:thres}
\end{equation}
The ansatz (\ref{e:thres}) transforms the equation for the electric field (\ref{e:eeqn}) into a linear, 
variable coefficient equation whose solution can be sought in the form
\begin{equation}
 u  =  u_{1}e^{-2iu_{1}^{2}z/q + i\sigma(\zeta)} + 
 u_{r} e^{-2iu_{1}^{2}z/q + i\sigma(\zeta)} ,
 \label{e:ures} 
\end{equation}
where $\sigma(\zeta)$ and $u_r(\zeta, z)$ are the phase correction due to the variable coefficient and  
the wavetrain amplitude, respectively.  To be consistent with the director (\ref{e:thres}) $u_{r} =
O(\epsilon^{2})$ as it is proportional to the jump height $u_{2} - u_{1}$.  Substituting (\ref{e:thres}) 
and (\ref{e:ures}) into the electric field equation (\ref{e:eeqn}) we have
\begin{eqnarray}
& &  i \frac{\partial u_{r}}{\partial z} - i(s_{+} - \sigma') \frac{\partial u_{r}}{\partial \zeta}
 + \frac{1}{2} \frac{\partial^{2} u_{r}}{\partial \zeta^{2}} 
 - \left( \frac{4\epsilon^{2}u_{1}}{q} A \hbox{sech}^{2}\beta \zeta -
 s_{+}\sigma' - \frac{i}{2} \sigma '' + \frac{1}{2} \sigma'^{2} \right) u_{r} \nonumber \\
 & & \mbox{} + \sigma' u_{1}s_{+} - \frac{1}{2} u_{1}\sigma'^{2} - 
 \frac{4\epsilon^{2}u_{1}^{2}}{q} A \hbox{sech}^{2} \beta \zeta = 0 .
 \label{e:reseqn}
\end{eqnarray}
We now choose the phase correction $\sigma(\zeta)$ so that the relation
\begin{equation}
 s_{+} \sigma ' - \frac{1}{2} \sigma'^{2} = \frac{4u_{1}\epsilon^{2}}{q}A \hbox{sech}^{2}\beta \zeta 
 \label{e:sigeqn}
\end{equation}
 is satisfied.
Then, on using (\ref{e:sigeqn}) to leading order in $\epsilon$, 
we obtain from (\ref{e:reseqn}) the equation for the variation of the wavetrain amplitude in the 
transition region as
\begin{equation}
 i \frac{\partial u_{r}}{\partial z} - i(s_{+} - \sigma') \frac{\partial u_{r}}
 {\partial \zeta} + \frac{1}{2} \frac{\partial^{2}u_{r}}{\partial \zeta^{2}}
 = 0 .
 \label{e:ureqn}
\end{equation}
In deriving this equation, we have noted that $\sigma''$ is higher order
in $\epsilon$ ( since $\beta \sim \epsilon/\sqrt{\nu}$, see (\ref{e:beta})).  

Using the numerical solution (see Fig.\ \ref{f:solnu1p5}(a)) and the soliton solution (\ref{e:kdvlead}) as a 
guide to the structure of the transition region we shall look for the solution of equation (\ref{e:ureqn}) 
for $u_r$ as fast (scaled as $O(1)$) oscillations with a slowly varying (scaled as $O(\beta^{-1})$) 
envelope.  
We then seek a WKB solution of the form
\begin{equation}
  u_{r} = W(X,Z) e^{i\psi(X,Z)/\beta} ,
 \label{e:wkb}
\end{equation}
where the slow variables are $X = \beta \zeta$ and $Z = \beta z$.  This WKB expansion is valid if 
$1/\sqrt{\nu} \ll \epsilon \ll \sqrt{\nu}$, which holds as $\nu$ is large.  The first inequality is
due to using the first two terms of the KdV expansion (\ref{e:thetakdv}) for the director (\ref{e:thres})
and the second is required for the validity of the WKB form (\ref{e:wkb}).  Substituting the WKB form 
(\ref{e:wkb}) into equation (\ref{e:ureqn}) gives the eikonal equation
\begin{equation}
 \frac{\partial \psi}{\partial Z} + \frac{1}{2}\left( \frac{\partial \psi}{\partial X}
 \right)^{2} - \left(s_{+} - \sigma '\right) \frac{\partial \psi}{\partial X} = 0
 \label{e:eik}
\end{equation}
and the transport equation
\begin{equation}
 \frac{\partial W}{\partial Z} + \left( \frac{\partial \psi}{\partial X}
 - s_{+} + \sigma'\right) \frac{\partial W}{\partial X} = - \frac{1}{2}
 \frac{\partial^{2} \psi}{\partial X^{2}} W .
 \label{e:trans}
\end{equation}

We note that the group and phase velocity argument gave that 
as the resonant wavetrain approaches the wavefront at $x = c_{g}z$,
it becomes a uniform wavetrain of wavenumber $k_{r}$ and frequency
$k_{r}^{2}/2 + 2\Theta_{1}$ \cite{nembore}.  We then find that the 
solution of the eikonal equation (\ref{e:eik}) is
\begin{equation}
 \psi = k_{r}X - \left( \frac{1}{2}k_{r}^{2} - s_{+}k_{r} \right) Z
 - \frac{4\epsilon^{2}u_{1}k_{r}}{qs_{+}(k_{r}-s_{+})\tilde{\beta}} A 
 \tanh X, \quad \tilde{\beta} = \sqrt{\nu} 
\beta = \frac{\epsilon \sqrt{A}}{\sqrt{2}(2q)^{1/4} \alpha} .
 \label{e:fsoln}
\end{equation}
We note that the phase correction (\ref{e:fsoln}) becomes infinite 
as $k_{r} \to s_{+}$.  This is expected as the group velocity of the front of 
the resonant wavetrain is $k_{r}$.  When the velocity of the lead soliton of the 
KdV DSW is greater than the group velocity, the wavetrain cannot propagate 
away from the DSW.  There is then no upstream resonant wavetrain, with 
only a small amplitude transient being present \cite{nembore}.

To solve the transport equation (\ref{e:trans}) the resonant wavetrain leading the 
KdV DSW must be matched to the intermediate shelf, so that $W=W_{0} = \epsilon^{2} = u_{2}-u_{1}$
at $X=0$ on noting the full solution (\ref{e:ures}) for $u$.  Then using the eikonal equation 
solution (\ref{e:fsoln}), the solution of the transport equation (\ref{e:trans}) is
\begin{equation}
 W = W_{0} \left[ 1 + \frac{2u_{1}\epsilon^{2}k_{r}A}{qs_{+}(k_{r}-s_{+})^{2}}\hbox{sech}^{2} X \right]
 \left[ 1 + \frac{2u_{1}\epsilon^{2}k_{r}A}{qs_{+}(k_{r}-s_{+})^{2}}\right]^{-1} .
 \label{e:w1}
\end{equation}
The height of the resonant wavetrain exponentially approaches the constant value 
\begin{equation}
W_c= W_{0}\left[ 1 + \frac{2u_{1}\epsilon^{2}k_{r}A}{qs_{+}(k_{r}-s_{+})^{2}}\right]^{-1}
\label{e:w0}
\end{equation}
as the front of the wavetrain at $x = c_{g}z$ is approached, so that the total height 
of the envelope of the resonant wavetrain in the region (ii) is given by
\begin{equation}
 a_{r} = u_{1} +  W_{0}\left[ 1 + \frac{2u_{1}\epsilon^{2}k_{r}A}{qs_{+}(k_{r}-s_{+})^{2}}\right]^{-1}.\label{e:ar}
\end{equation}

Finally, we describe region (iii) of the resonant wavetrain which brings it
down to the initial level $u_{1}$ (see Figure \ref{f:solnu1p5}(a)).  In the region of this front, as for 
the uniform middle region (ii), we approximate $\theta$ by $\theta = \Theta_{1} = |u_{1}|^{2}/q$, so that 
the linear equation (\ref{e:eeqnwave}) holds.  If we use a moving coordinate 
$\zeta_{g} = x - c_{g}z$ moving with the velocity of the front, the electric field is governed by 
\begin{equation}
 i \frac{\partial u}{\partial z} - ic_{g} \frac{\partial u}{\partial \zeta_{g}} 
 + \frac{1}{2} \frac{\partial^{2} u}{\partial \zeta_{g}^{2}} - 2\Theta_{1}u = 0.
 \label{e:ufront}
\end{equation}
To match with the initial level ahead, we seek a solution of the form
\begin{equation}
 u = u_{1}e^{-2i\Theta_{1}z} + u_{f}e^{-2i\Theta_{1}z},
 \label{e:uf1}
\end{equation}
so that $u_{f}$ is the solution of
\begin{equation}
 i \frac{\partial u_{f}}{\partial z} - ic_{g} \frac{\partial u_{f}}{\partial \zeta_{g}} 
 + \frac{1}{2} \frac{\partial^{2} u_{f}}{\partial \zeta_{g}^{2}} = 0.
 \label{e:ufeqn}
\end{equation}
To match with the uniform wavetrain behind, we have the boundary condition $|u_{f}| = a_{r} - u_{1}$
at $\zeta_{g} = 0$.  The linear equation (\ref{e:ufeqn}) can be solved using Laplace transforms
to give the Fresnel integral solution
\begin{equation}
 u_{f} = \frac{2\left( a_{r} - u_{1} \right)}{\sqrt{\pi}} e^{i\left(c_{g}\zeta_{g} + \frac{1}{2}c_{g}^{2}z
 - \pi/4\right)} \int_{\frac{\zeta_{g}}{\sqrt{2z}}}^{\infty} e^{it^{2}} \: dt.
 \label{e:fresnel}
\end{equation}

\begin{figure}
\centering
\includegraphics[width=0.45\textwidth,angle=270]{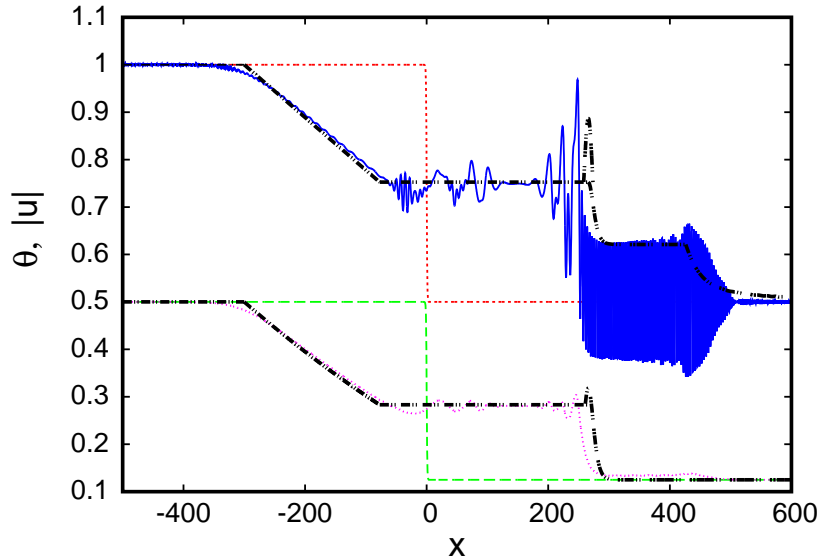}
\caption{Comparison between numerical solution of nematic equations (\ref{e:eeqn}) and (\ref{e:direqn})
and the modulation theory solution of Sections \ref{s:riemann}\ref{s:nondisp}, \ref{s:riemann}\ref{s:kdvbore}
and \ref{s:riemann}\ref{s:wkb} for $u_{3}=1.0$, $u_{1}=0.5$, $q=2$ and $\nu = 200$.  Initial condition for 
$|u|$ ($z=0$):  red (short dashed) line; initial condition for $\theta$ ($z=0$):  green (long dash) line; numerical 
solution for $|u|$ at $z=300$:  blue (solid) line; numerical solution for $\theta$ at $z=300$: pink 
(dotted) line; modulation solution:  black (dot-dot-dash) line. Only the lead soliton of the modulation 
theory DSW solution is shown.}
\label{f:compu1p5}
\end{figure}

\section{Comparison with numerical solutions}
\label{s:num}

In this section, full numerical solutions of the nematic equations (\ref{e:eeqn}) and (\ref{e:direqn})
will be compared with the modulation theory solutions of Sections \ref{s:riemann} \ref{s:nondisp}, 
\ref{s:kdvbore} and \ref{s:wkb}.  The numerical solution of the electric field equation (\ref{e:eeqn}), 
which is of NLS-type, was obtained using the pseudo-spectral method of Fornberg and Whitham \cite{bengt}, modified
to improve its accuracy and stability \cite{if}, but without the boundary damper due to the non-zero boundary
conditions.  These improvements include using a 4th order Runge-Kutta
scheme to propagate forward in $z$, resulting in higher accuracy, in Fourier space, rather than in real space,
resulting in improved stability \cite{if}.  The numerical solution of the linear director equation (\ref{e:direqn}) 
was obtained using a spectral method \cite{numrec}.  This numerical scheme is discussed in \cite{benthesis}.
For the numerical solutions of this work $32768$ points were used for the FFT with a $x$ domain of length $8192.0$
and a $z$ step of $\Delta z = 0.002$.  The $x$ domain was chosen long enough so that the waves at the numerical 
boundaries generated by periodicity were far from the region of interest.  Finally, the initial condition
(\ref{e:ic}) was smoothed using $\tanh x/W$ to avoid spurious numerical effects due to large $x$ derivatives, with
$W=1$ found to be suitable.  

Figure \ref{f:compu1p5} shows a comparison between the numerical solution of the nematic equations
(\ref{e:eeqn}) and (\ref{e:direqn}) for $u_{3} = 1.0$ and $u_{1}=0.5$ at $z=300$ for $q=2$ and $\nu=200$.
For clarity, in these figures only the upper envelope of the resonant wavetrain (\ref{e:ures}), (\ref{e:wkb}) 
and (\ref{e:w1}) and the upper envelope of the Fresnel front (\ref{e:fresnel}) are shown.   
It can be seen that there is very good agreement in general between the numerical solution for 
the electric field $|u|$ and the modulation theory solution of Sections 5\ref{s:nondisp}, 5\ref{s:kdvbore} and 
5\ref{s:wkb}.  In particular, there is excellent agreement for the position of the lead soliton of the 
DSW, which is the same as that of the trailing edge of the resonant wavetrain.  
This is in contrast to the result of previous work \cite{nembore} in which this position 
was determined by the velocity $s_{+}=\sqrt{2/q}u_{3}$  of the lead soliton of the 
standard KdV DSW solution \cite{gur,bengt}, resulting in the DSW leading edge being at $x=300$ 
for the parameters of Figure \ref{f:compu1p5}, noting that the numerical position is 
$x=247.5$.  It is then clear that the shock velocity (\ref{e:v2sp}) determined
from the shock jump conditions for the non-dispersive ``shallow water'' equations 
(\ref{e:massnd})--(\ref{e:thmnd}) and giving $x=265.9$ for the 
lead soliton at $z=300$ yields much better agreement with the numerical solution for 
the position of the leading edge of the DSW than the velocity determined by the 
KdV DSW solution.  The differing length scales of the KdV DSW ($O(\sqrt{\nu})$) and the resonant
wavetrain ($O(1)$) can be clearly seen.  The major disagreement is that the front of the numerical 
resonant wavetrain has more structure than the linear Fresnel integral solution of Section 
5\ref{s:wkb}.  However, the Fresnel integral solution gives the correct spatial extent of 
the transient front of the resonant wavetrain.  Furthermore, if the Fresnel integral 
solution is shifted so that it starts ahead of the rise in the numerical front, it is 
in very good agreement with the numerical front.   

The other noticeable disagreement between the numerical and analytical solutions is the 
amplitude of the lead soliton of the DSW.  The amplitude of the DSW in the electric field $u$ is  
generally under-predicted by the KdV theory of Section 5\ref{s:kdvbore}, which was based on the classical shock 
speed (\ref{e:v2sp}).  However, this approximation yields good agreement for the DSW in the director 
$\theta$, given in the KdV approximation by Eq.\ (\ref{e:thres}).  This is in contrast to the results of 
\cite{nembore} for which the standard DSW solution of the KdV equation was used to determine the 
DSW on the intermediate level $u_{2}$.  The results of \cite{nembore} strongly over-predicted the height 
of the bore in the director, this major discrepancy being fixed in the present theory.
Finally, it can be seen that under the resonant wavetrain there is a slight rise in the 
director above $\theta = \Theta_{1}$ due to $O(\nu^{-1})$ corrections in the asymptotic 
expansions.  These higher order corrections will be dealt with in future work based on
a full description of a resonantly radiating DSW.

\begin{figure}
\centering
\includegraphics[width=0.33\textwidth,angle=270]{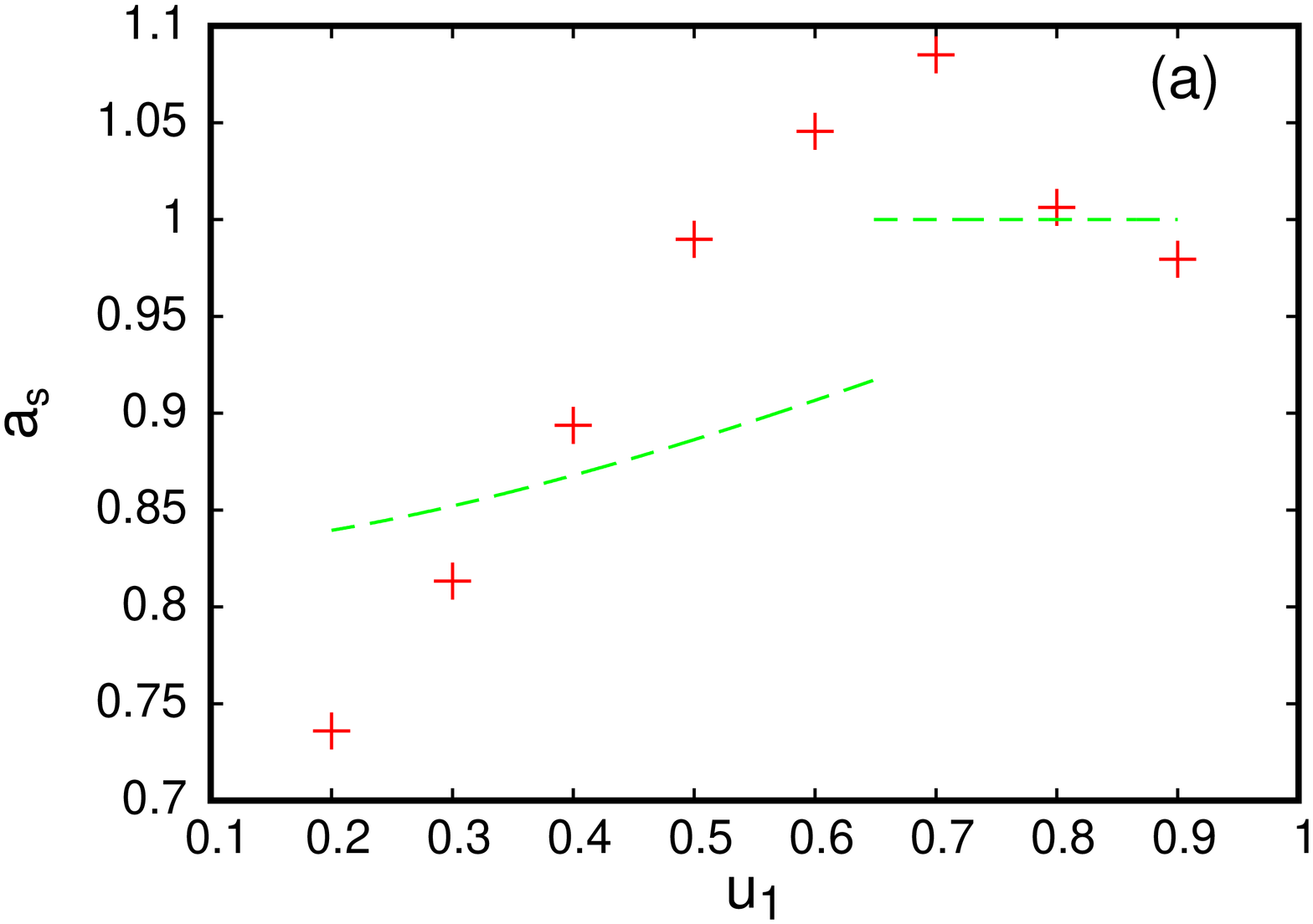}
\includegraphics[width=0.33\textwidth,angle=270]{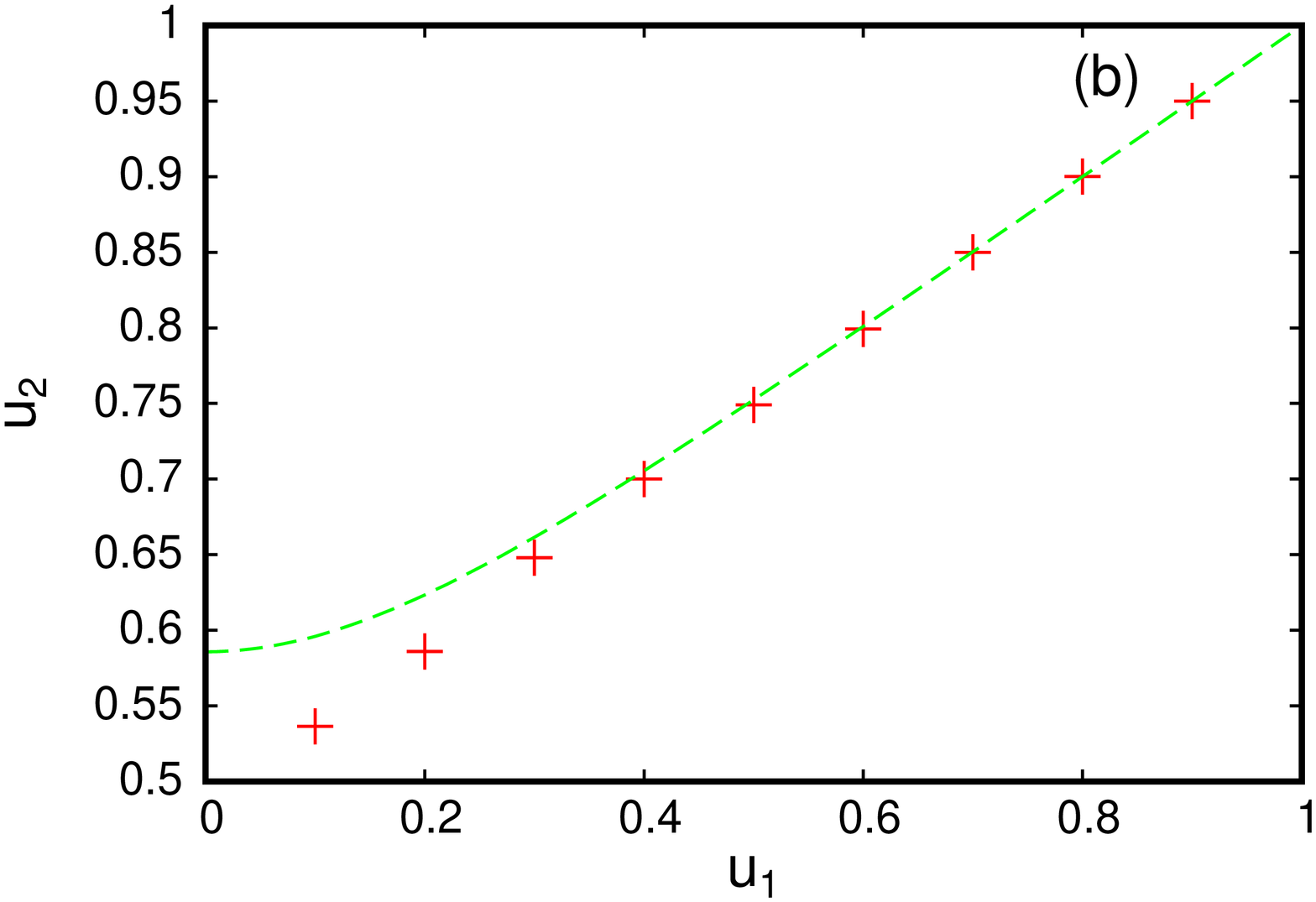}
\includegraphics[width=0.33\textwidth,angle=270]{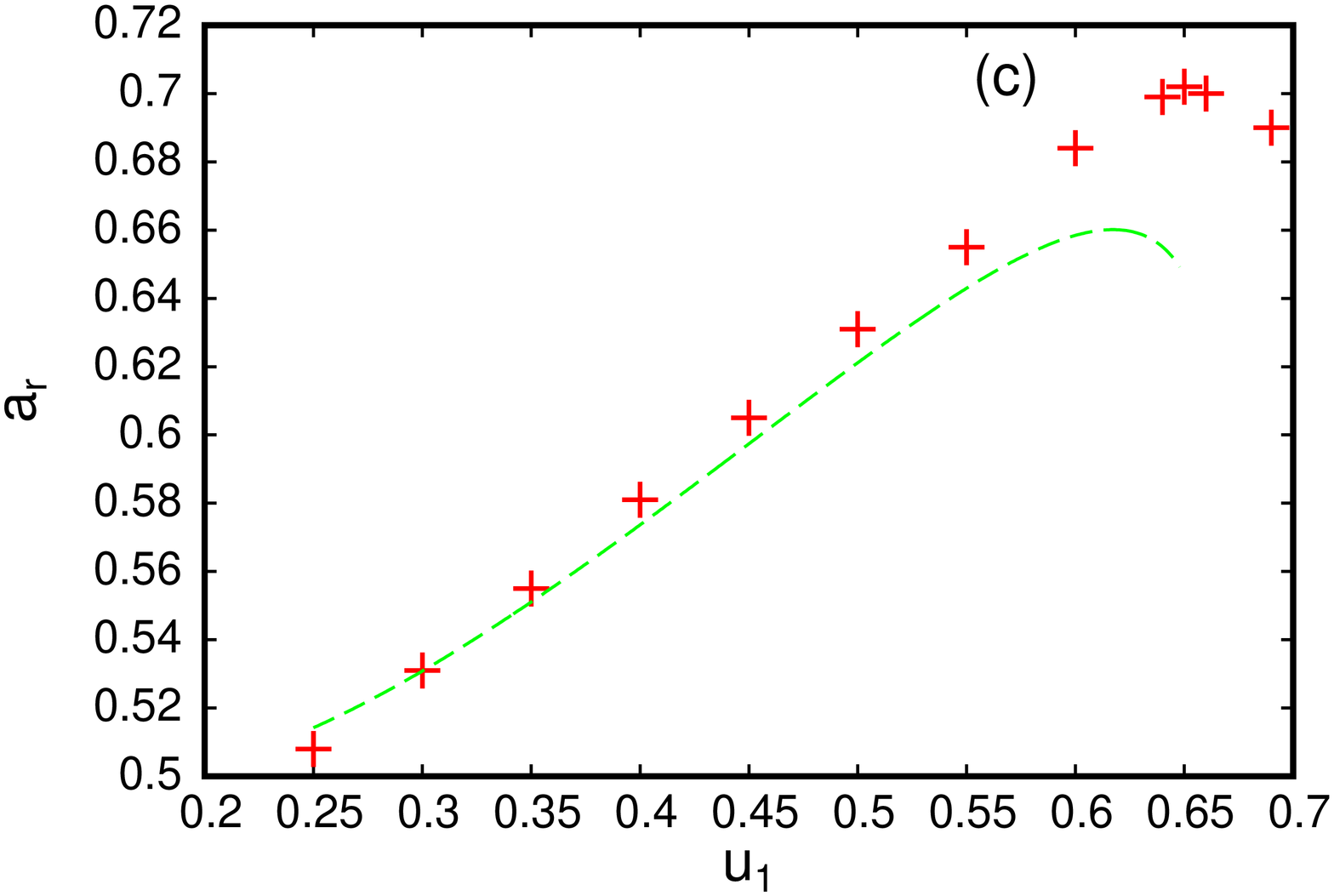}
\includegraphics[width=0.33\textwidth,angle=270]{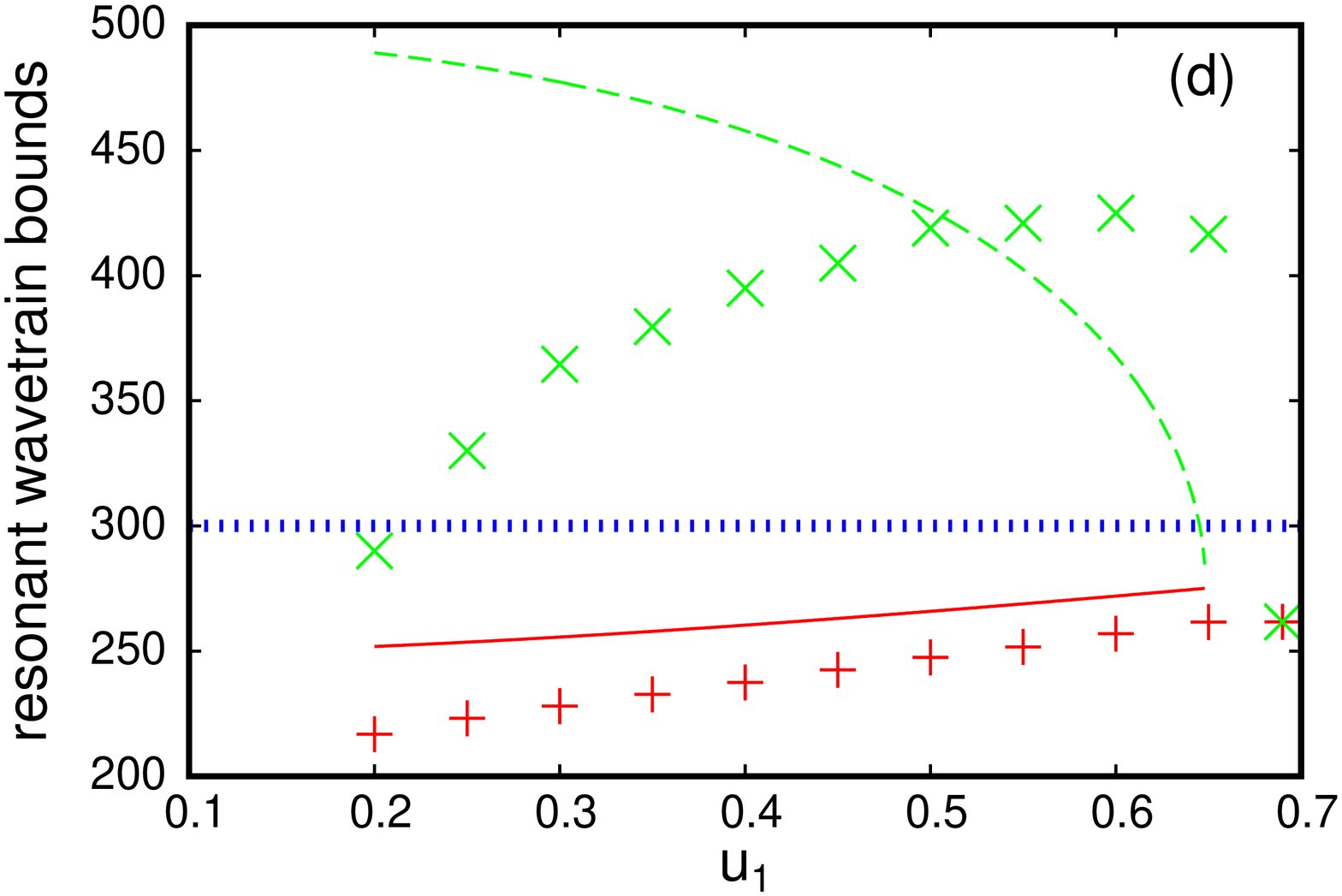}
\caption{(a)  Height $a_{s}$ of lead soliton of the DSW.  Numerical solution of 
(\ref{e:eeqn}) and (\ref{e:direqn}):  red pluses; analytical  solution $a_s=u_1+\epsilon^2 A$ (\ref{e:kdvboreamp})
and (\ref{e:kdvboreampnores}):  green (dashed) line. (b) Height $u_{2}$ of intermediate
shelf.  Numerical solution of (\ref{e:eeqn}) and (\ref{e:direqn}):  red pluses; 
modulation solution (\ref{e:u2expr}): green (dashed) line.  (c) Height $a_r$ of resonant 
wavetrain as a function of the upstream level $u_{1}$.  Numerical solution of 
(\ref{e:eeqn}) and (\ref{e:direqn}):  red pluses; WKB solution (\ref{e:ar}):  green 
(dashed) line.  (d) Comparison for leading and trailing edges of resonant wavetrain at 
$z=300$.  Numerical trailing edge: red pluses;  trailing edge $x_-=s_+z$ given by the 
classical shock speed $s_+$ (\ref{e:phasegroup}): red (solid) line; numerical leading 
edge:  green crosses; leading edge $x_+=c_g z$ defined by the group velocity 
(\ref{e:group}): green (dashed) line; trailing edge of \cite{nembore} given by the 
soliton speed (\ref{e:kdvboreampnores}) in the standard modulation solution for the 
KdV DSW:  blue (dotted) line.   The other parameter values are $u_{3}=1$, $q=2$ and 
$\nu = 200$.}
\label{f:resamp}
\end{figure}

The agreement between the modulation theory and numerical solutions is further 
quantified in Figure \ref{f:resamp}.  Figure \ref{f:resamp}(a) shows a comparison of the 
height (background plus amplitude) of the lead soliton of the DSW as given by numerical 
solutions and the modulation solution ($a_s=\epsilon ^{2}A + u_{1}$), using  
(\ref{e:kdvboreamp}) for the amplitude below the cut-off (\ref{e:resbound}) and  
(\ref{e:kdvboreampnores}) above.  The choice of the total height rather than amplitude 
for the comparisons is due to the soliton background being not clearly defined in the 
numerical solutions (see Fig.\ \ref{f:solnu1p5}(a)).  Furthermore, the amplitude of the
lead wave oscillates slightly due to its interaction with the resonant wavetrain, so the 
figure shows the average amplitude.  The numerical solution clearly shows 
the predicted different DSW behaviours above and below the resonant wave cut-off, which 
was not predicted in \cite{nembore} for which the height was the constant value 
(\ref{e:kdvboreampnores}) for the whole range of $u_1$.  
Thus, one can see that the KdV soliton height based on the classical shock wave speed
is in broad agreement with the numerical values.  
The appropriateness of using the classical shock wave velocity to determine the intermediate 
shelf height (\ref{e:u2expr}) is quantified in Figure \ref{f:resamp}(b).  It can be seen that (\ref{e:u2expr})
is in excellent agreement with the numerical height, except for a slight discrepancy
as $u_{1} \to 0$.  This is due to the intermediate shelf disappearing as the 
dam break solution for $u_{1}=0$ is approached \cite{nembore}.  

Figure \ref{f:resamp}(c) shows a comparison between the height of the resonant wavetrain obtained 
numerically and the modulation solution height (\ref{e:ar}).  There is excellent agreement between these 
heights, except towards the cut-off near $u_{1}=0.7$.  This is due to the discrepancy 
between the numerically found cut-off and the modulation theory prediction.  

Finally, Figure \ref{f:resamp}(d) shows a comparison for the leading and trailing edges of the resonant 
wavetrain.  It can be seen that there is excellent agreement for the position of the 
trailing edge, even up to the cut-off.  Previous work \cite{nembore} predicted the 
constant (i.e.\ independent of $u_1$) velocity (\ref{e:kdvboreampnores}) for the 
trailing edge which was defined by the value $u_3$ alone.  Figure \ref{f:resamp}(d) 
clearly shows that the present theory based on the classical shock speed for the leading edge of the 
DSW is in much better agreement.  The agreement for the leading edge is reasonable above 
$u_{1}=0.5$ as the cut-off is approached, but is poor as $u_{1}$ decreases.  There are a number of reasons 
for this.  The position of the trailing edge is clearly defined by the peak of the lead soliton
of the KdV DSW.  While the theory of Section 5\ref{s:wkb} predicts a precise location
for the leading edge of the resonant wavetrain, it can be seen from Figs.\ \ref{f:solnu1p5}(a) and 
\ref{f:compu1p5} that this is not the case for the numerical solution.  There is no clean boundary 
between the resonant wavetrain and its front.  There is an extended transition between the two.  For 
the comparison of Figure \ref{f:resamp}(d) the start of the hump before the monotonic decrease of 
the front was chosen as the leading edge position of the wavetrain.  
Finally, the present modulation solution underpredicts the cut-off point for the resonant wavetrain
(at $u_{1}=0.648$ compared with the numerical value $0.69$ for the parameter values of Fig.\ \ref{f:resamp}), 
leading to the disagreement as the cut-off is approached.

Part of the reasons for these discrepancies is due to the major simplifying assumption adopted in 
the present theory, according to which the generation of the wavetrain is dominated by a single resonance with 
the lead soliton of the DSW.  A more advanced theory including internal resonances with the other 
components of the DSW is needed to achieve better agreement with numerical solutions.  

\begin{figure}
\centering
\includegraphics[width=0.33\textwidth,angle=270]{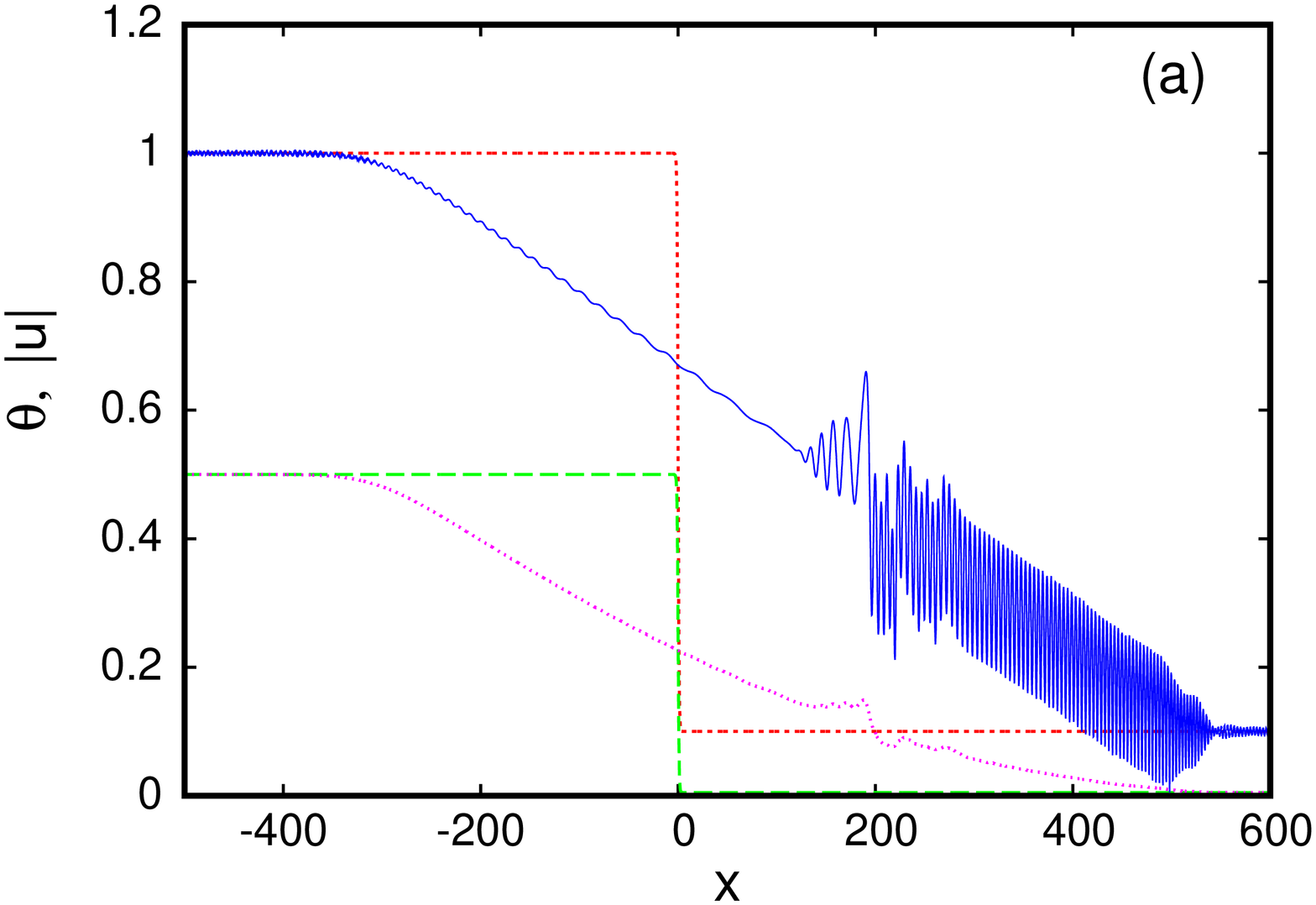}
\includegraphics[width=0.33\textwidth,angle=270]{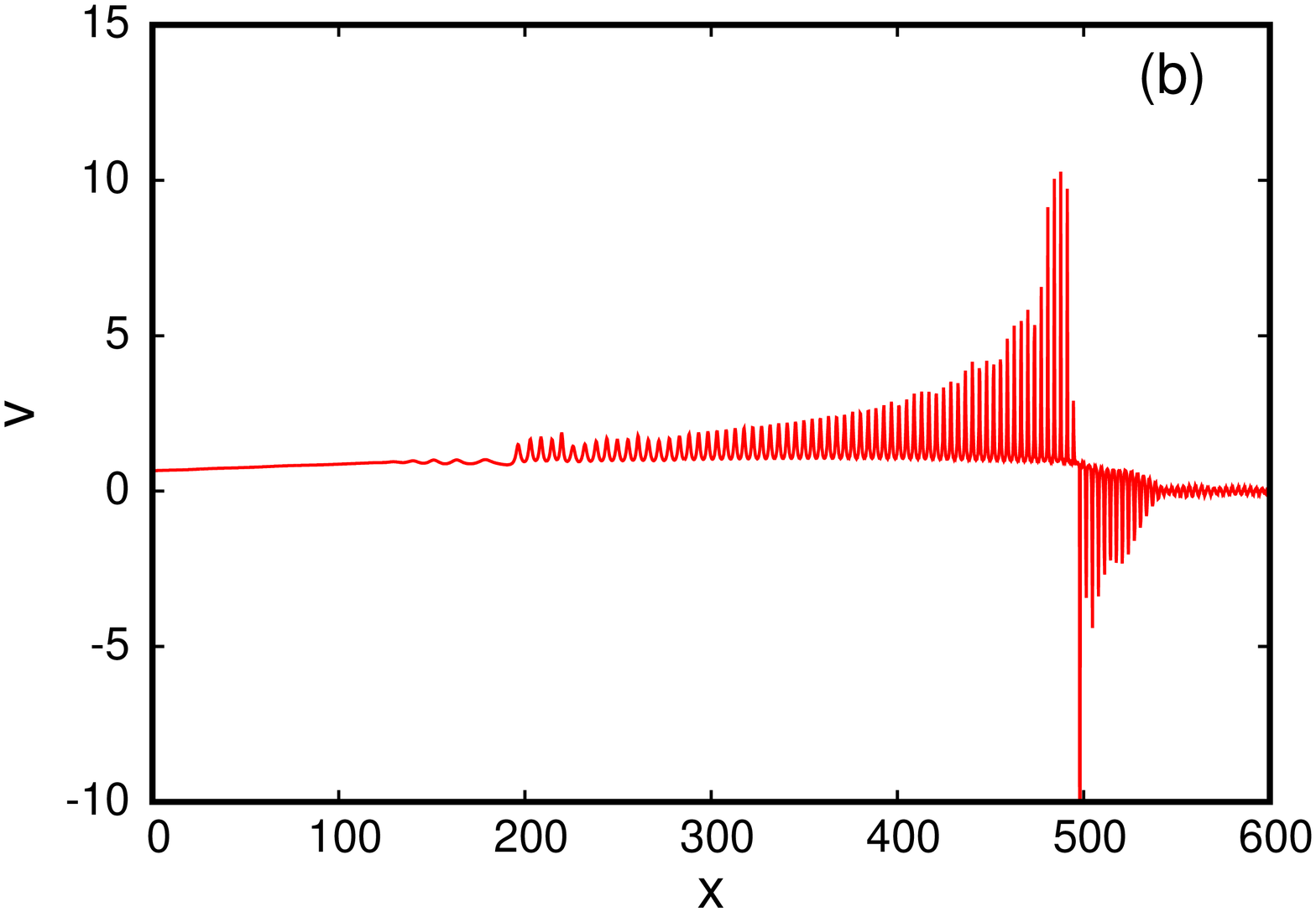}
\caption{Numerical solution of the nematic equations (\ref{e:eeqn}) and (\ref{e:direqn})
for $u_{3}=1.0$, $u_{1}=0.1$, $q=2$ and $\nu = 200$.  (a) Initial condition for $|u|$ ($z=0$):  red 
(short dashed) line; initial condition for $\theta$ ($z=0$):  green (long dash) line; numerical solution 
for $|u|$ at $z=300$:  blue (solid) line; numerical solution for $\theta$ at $z=300$: pink (dotted) line, 
(b) phase $v$ at $z=300$.}
\label{f:solnu1p1}
\end{figure}

There is one more feature of the resonant wavetrain which complicates its analysis for large initial 
jumps.  As the initial level ahead $u_{1}$ decreases the electric field $u$ eventually vanishes at 
a point, termed the vacuum point \cite{gennady}.  For sufficiently large initial jumps the vacuum point occurs 
within the resonant wavetrain, so that the lower envelope becomes non-monotone.  It was shown in \cite{gennady} 
that for the defocusing NLS DSW there is a singularity in the phase $v$ at the vacuum point itself.
Although the resonant wavetrain for the nematic system (\ref{e:eeqn}) and (\ref{e:direqn}) is asymptotically 
described by the linear equation (\ref{e:eeqnwave}) rather than the defocusing NLS equation, numerical simulations 
show that the vacuum point in the wavetrain has qualitatively similar properties to the vacuum point arising in the 
large amplitude NLS DSW \cite{gennady}.  In particular, such a DSW has a non-monotone lower envelope 
(see Figure \ref{f:solnu1p1}(a)) and exhibits a phase singularity at the vacuum point, see Figure 
\ref{f:solnu1p1}(b).  The WKB solution of Section 5\ref{s:wkb} gives that the lower envelope of the resonant 
wavetrain has height 
\begin{equation}
 a_{l} = u_1 - W_c=u_{1} - W_{0}\left[ 1 + \frac{2u_{1}\epsilon^{2}k_{r}A}{qs_{+}(k_{r}-s_{+})^{2}}\right]^{-1} .
\label{e:al}
\end{equation}
For $u_{3}=1.0$, $\nu=200$ and $q=2$, it is found that $a_{l}$ vanishes when $u_{1}=0.2416$.
Numerical solutions of the nematic equations (\ref{e:eeqn}) and (\ref{e:direqn}) show that 
for these parameter values a vacuum point first occurs when $u_{1}=0.22$.  A full analysis
of the solution after the vacuum point is reached is beyond the scope of this paper.  Full Whitham
modulation equations would be required for a proper analysis after the vacuum point \cite{gennady}.

\section{Conclusions}

The Riemann problem for the equations governing the propagation of a coherent 
optical beam in a defocusing nematic liquid crystal has been studied.  It was found that in 
the highly nonlocal limit the DSW, which comprises a major part of the Riemann problem 
solution, is drastically different to the DSW solution of the defocusing NLS equation, to which 
the nematic equations reduce in the small nonlocality limit, that is $\nu \to 0$.  There are two major 
differences: (i) the nematic DSW is of positive polarity with a bright soliton at its leading edge; (ii)  it 
is preceded by a short wavelength resonant wavetrain.  To clarify this structure, 
it was shown that in the limit of small deviations from a background, the nematic
equations reduce to a KdV equation with a fifth order derivative, the Kawahara equation.
This fifth order KdV equation is known to have a resonance between its solitary wave 
solution and linear radiation.  The present work shows that this resonance extends to a 
resonance between the DSW and linear radiation.   A modulation theory was developed to derive 
solutions for the resonant wavetrain and its front.  In contrast to previous work \cite{nembore}, it was 
found that the leading edge of the DSW was determined by the classical shock jump condition,
which is non-standard for DSWs \cite{el2}.  Excellent agreement was found between 
the major part of the modulation theory solution and full numerical solutions of the 
nematic equations.  However, there are some discrepancies.  Part of the observed discrepancies can be 
addressed by applying a more complete modulation theory for the DSW description.
When the higher order dispersion term is a small perturbation, as in the Kawahara equation 
(\ref{e:kdv5norm}) with $\gamma \ll 1$, Whitham theory for perturbed integrable 
equations \cite{kamch2004,kamch2015} provides an appropriate analytical framework for the 
description of the DSW evolution.

The present work leaves open a number of issues.  As already discussed, resonances between 
DSWs and radiation is an issue which has received little attention to date with 
the theory and solution methods only starting to be developed \cite{bakholdin, trillo1}, in 
contrast to the corresponding resonant interaction between solitary waves and radiation 
(see e.g.\ \cite{grim93, radiating_sol1, radiating_sol2, kivshar, supercontinuum}).  
As the nematic equations are generic and similar equations arise in a number of fields, 
this resonant interaction deserves in depth treatment.   A proper analysis of 
possible resonances between DSWs and radiation is an open question which deserves 
further treatment.  This will be the subject of further work.

\section*{Acknowledgement}

GE and NFS were supported by the London Mathematical Society under the Research in
Pairs Grant 41421.  GE was also partially supported by the Royal Society International Exchanges 
Scheme Grant IE131353.

\bibliographystyle{vancouver}

\end{document}